\documentstyle[12pt]{article}

\catcode`\@=11
\def\marginnote#1{}
\newcount\hour
\newcount\minute
\newtoks\amorpm
\hour=\time\divide\hour by60
\minute=\time{\multiply\hour by60 \global\advance\minute by-\hour}
\edef\standardtime{{\ifnum\hour<12 \global\amorpm={am}%
        \else\global\amorpm={pm}\advance\hour by-12 \fi
        \ifnum\hour=0 \hour=12 \fi
        \number\hour:\ifnum\minute<10 0\fi\number\minute\the\amorpm}}
\edef\militarytime{\number\hour:\ifnum\minute<10 0\fi\number\minute}
\def\draftlabel#1{{\@bsphack\if@filesw {\let\thepage\relax
   \xdef\@gtempa{\write\@auxout{\string
      \newlabel{#1}{{\@currentlabel}{\thepage}}}}}\@gtempa
   \if@nobreak \ifvmode\nobreak\fi\fi\fi\@esphack}
        \gdef\@eqnlabel{#1}}
\def\@eqnlabel{}
\def\@vacuum{}
\def\draftmarginnote#1{\marginpar{\raggedright\scriptsize\tt#1}}
\def\draft{\oddsidemargin -.5truein
        \def\@oddfoot{\sl preliminary draft \hfil
        \rm\thepage\hfil\sl\today\quad\militarytime}
        \let\@evenfoot\@oddfoot \overfullrule 3pt
        \let\label=\draftlabel
        \let\marginnote=\draftmarginnote
   \def\@eqnnum{(\theequation)\rlap{\kern\marginparsep\tt\@eqnlabel}%
\global\let\@eqnlabel\@vacuum}  }

\def\preprint{\twocolumn\sloppy\flushbottom\parindent 1em
        \leftmargini 2em\leftmarginv .5em\leftmarginvi .5em
        \oddsidemargin -.5in    \evensidemargin -.5in
        \columnsep 15mm \footheight 0pt
        \textwidth 250mmin      \topmargin  -.4in
        \headheight 12pt \topskip .4in
        \textheight 175mm
        \footskip 0pt
        \def\@oddhead{\thepage\hfil\addtocounter{page}{1}\thepage}
        \let\@evenhead\@oddhead \def\@oddfoot{} \def\@evenfoot{} }

\def\titlepage{\@restonecolfalse\if@twocolumn\@restonecoltrue\onecolumn
     \else \newpage \fi \thispagestyle{empty}\c@page\z@
        \def\thefootnote{\fnsymbol{footnote}} }

\def\endtitlepage{\if@restonecol\twocolumn \else  \fi
        \def\thefootnote{\arabic{footnote}}
        \setcounter{footnote}{0}}  

\catcode`@=12
\relax
\def\bea{\begin{array}}
\def\bem{\begin{displaymath}}
\def\beq{\begin{equation}}
\def\eea{\end{array}}
\def\eem{\end{displaymath}}
\def\eeq{\end{equation}}

\def\Im{\mathop{\rm Im}}
\def\NP#1#2#3{Nucl. Phys. \underline{#1} (19#2) #3}
\def\ov{\overline}

\def\PL#1#2#3{Phys. Lett. \underline{#1} (19#2) #3}
\def\PR#1#2#3{Phys. Rev. \underline{#1} (19#2) #3}
\def\PRL#1#2#3{Phys. Rev. Lett. \underline{#1} (19#2) #3}
\def\Re{\mathop{\rm Re}}

\relax
\newcommand{\be}{\begin{equation}}
\newcommand{\en}{\end{equation}}
\newcommand{\ba}{\begin{eqnarray}}
\newcommand{\ea}{\end{eqnarray}}
\newcommand{\ee}{\end{equation}}
\def\alp{\alpha^\prime}

\def\crbig{\\\noalign{\vspace {3mm}}}

\def\slash #1{{\not \hspace{-.5mm}#1}}
\relax

\begin{document}
\topmargin-2.4cm
\renewcommand{\theequation}{\thesection.\arabic{equation}}
%
%
%
%
\begin{titlepage}
\begin{flushright}
CERN--TH/99-20\\
LPTENS-98/48\\
CPTH-S706.1298\\
NEIP--98--021 \\
hep--th/9902032 \\
February 1999
\end{flushright}
\vspace{.3cm}
\begin{center}{\Large\bf
Non-perturbative Temperature Instabilities in {\bf $N=4$}
Strings$^{\star}$ }
\vskip .15in
{\bf I. Antoniadis$^{\,a}$,  J.-P. Derendinger$^{\, b}$
and C. Kounnas$^{\,c,{\dagger\dagger}}$}
\vskip .3cm
{\normalsize\sl
$^a$Centre de Physique Th{\'e}orique, Ecole Polytechnique,$^\dagger$
\\
F-91128 Palaiseau, France}\\ [3mm]
{\normalsize\sl $^b$Institut de Physique, Universit\'e de
Neuch\^atel,\\
Breguet 1, CH-2000 Neuch\^atel, Switzerland}\\[3mm]
{\normalsize\sl $^c$Theory Division, CERN, 1211 Geneva 23,
Switzerland}
\end{center}
\vskip .3cm
\begin{center}
{\bf Abstract}
\end{center}
\begin{quote}
We derive a universal thermal effective potential, which describes
all possible high-temperature instabilities of the known $N=4$
superstrings, using the properties of gauged $N=4$ supergravity.
These instabilities are due to three non-perturbative thermal dyonic
modes, which become tachyonic in a region of the thermal moduli
space. The latter is described by three moduli, $s,t,u$, which are
common to all non-perturbative dual-equivalent strings with $N=4$
supersymmetry in five dimensions: the heterotic on $T^4\times S^1$,
the type IIA on $K_3\times S^1$, the type IIB on $K_3\times S^1$ and
the type I on $T^4\times S^1$. The non-perturbative instabilities are
analysed. These strings undergo a high-temperature transition to a
new phase in which five-branes condense. This phase is described in
detail, using both the effective supergravity and non-critical string
theory in six dimensions. In the new phase, supersymmetry is
perturbatively restored but broken at the non-perturbative level. In
the infinite-temperature limit the theory is topological with an
$N=2$ supersymmetry based on a topologically non-trivial
hyper-K\"ahler manifold.
\end{quote}

\vspace{.4cm}
\begin{flushleft}
\rule{8.1cm}{0.2mm}\\
$^{\star}$
{\small Research supported in part by
the EEC under the TMR contracts ERBFMRX-CT96-0090 and
ERBFMRX-CT96-0045, and by the Swiss National Science
Foundation and the Swiss Office for Education and Science.} \\
$^{\dagger}$
{\small Unit\'e mixte du CNRS UMR 7644.} \\
$^{\dagger\dagger}$ {\small On leave from Ecole Normale
Sup\'erieure, 24 rue Lhomond, F-75231 Paris Cedex 05,
France.}
\end{flushleft}

\end{titlepage}
\setcounter{footnote}{0}
\setcounter{page}{0}
\setlength{\baselineskip}{.7cm}
\newpage
%
%
\section{Introduction}\label{secintro}
\setcounter{equation}{0}

In a physical system where the density of states grows
exponentially with the energy,
\beq
\label{density}
\rho(E)\sim E^{-k}e^{bE}\, ,
\eeq
there is a critical temperature, $\beta^{-1}\equiv T= T_H=b^{-1}$, at
which various thermodynamical quantities diverge \cite{H}. In
particular, the partition function $Z$ and the mean energy $U$
develop power pole singularities:
\beq
\label{Z}
\begin{array}{rcl}
Z(\beta) &=&\displaystyle{\int dE\,\rho(E) e^{-\beta E}
\sim {1\over{(\beta-b)}^{(k-1)}}\, },\crbig
U(\beta) &=& \displaystyle{ -{\partial\over\partial\beta}\ln Z \sim
(k-1){1\over\beta-b}+{\rm regular}\, .}
\end{array}
\eeq
An alternative interpretation of the mean energy pole singularity
follows from the identification of the temperature with the inverse
radius of a compactified Euclidean time on $S^1$. In this
representation, the partition function is given by the (super)trace
over the thermal spectrum of the theory in one dimension less:
\beq
\ln Z={\rm Str}\ln{\cal M}(\beta)\, .
\eeq
The pole singularity is then a manifestation of a thermal state that
becomes massless at the critical temperature. Thus, the knowledge of
the thermal spectrum of the theory ${\cal M}(\beta)$, as a function
of the $S^1$ radius $R=\beta/2\pi$, determines the critical
temperature \cite{S}--\cite{AK}.

Perturbative string theory provides an example of an exponentially
growing density of states, with $k$ in Eq.(\ref{density}) equal to
the dimension of space-time, and the exponent $b^{-1}$, and thus the
Hagedorn temperature, given as a theory-dependent constant in terms
of the string scale $(\alpha^\prime)^{-1/2}$ \cite{A,AEN,AxK,KR,AK}.
In the picture where temperature is regarded as a compactification on
a circle of radius $R$, one can exactly construct the partition
function $Z(R)$ and identify the state that becomes tachyonic at the
critical temperature \cite{AW,AK}. This state has necessarily a
non-zero winding number $n$, as perturbative quantum field theory is
not able to generate a similar critical behaviour. A detailed
discussion of this phenomenon in perturbative string theory will be
the subject of the next section.

It is interesting that one can go a long way into the discussion of
thermal instabilities due to non-perturbative string states, and then
also, of non-perturbative field theory states. The first observation
is that, in $N_4=4$ supersymmetric strings\footnote{$N_D$ is the
number of $D$-dimensional supersymmetries.} (or $N_6=2$), the {\it
perturbative} string states becoming tachyonic above the Hagedorn
temperature are (thermally-shifted) BPS states that preserve half of
the supersymmetries ($N_4=2$ or $N_6=1$). In these theories, the
masses of {\it non-perturbative} BPS states are also known from
$N_4=4$ supersymmetry \cite{N=4BPS,DVV,KK} and one can identify among
them the states that are able to induce a thermal instability and the
critical temperature at which they become tachyonic. We will develop
this argument in Section \ref{secdual}, using heterotic--type II
duality \cite{HT}--\cite{sixdimdual}.

Notice that considering only those $N_4=4$ BPS states preserving
$N_4=2$ supersymmetries (1/2-BPS) is certainly sufficient to study
thermal instabilities in, for instance, non-perturbative heterotic
strings in space-time dimensions $D\ge6$. BPS states preserving less
supersymmetries only arise in lower dimensions (1/4-BPS, with
$N_4=1$, for $D\le5$ or 1/8-BPS, with $N_2=1$, for $D\le3$). In
dimension six or higher, it is expected that thermal instabilities
due to $N_4=2$ BPS states are similar to those of perturbative
winding states, with an entropy growing linearly with the mass. This
statement, as we will see in the following sections, can be checked
in six dimensions, since heterotic--type II duality allows a
non-perturbative behaviour on one side to be turned to a perturbative
one on the dual side. In dimensions lower than six, the 1/4- or
1/8-BPS states singularities have to correspond to an entropy growing
with the mass faster than linearly. This statement is supported by
the behaviour of black hole entropy in four and five dimensions,
which follows the area law \cite{areaBPS}. In this case, the
temperature is fixed and the canonical ensemble does not exist
\cite{AxK}. However, we stress that the analysis given in this paper
is fully general in dimensions greater than (or equal to) six, while
it only applies to 1/2-BPS states in lower dimensions.

With this limitation in mind, the main result of this paper is a
computation of the exact effective potential for all the potentially
tachyonic states, as a function of the
(universal\footnote{Invariant under string dualities.})
temperature. It reproduces all known Hagedorn temperatures for
heterotic and type II strings in the appropriate limits. The exact
potential has a global minimum in a domain of the six-dimensional
string coupling that includes the perturbative heterotic regime. In
this phase, supersymmetry is perturbatively restored, the temperature
is fixed, $T^{-1} = \pi\sqrt{2\alpha^\prime_H}$ and space-time
geometry is that of the heterotic or type IIA five-brane. More
precisely, the system loses four units of central charge. It
describes a non-critical string in six dimensions with massless
thermal excitations extending the concept of particles with infinite
correlation length in finite-temperature field theory with a
second-order phase transition. On the type II side, this phase is
characterized by a condensation of five-branes.

The present paper is organized as follows. In Section \ref{secpert},
we recall the aspects of perturbative strings at finite temperature,
which will be used in the non-perturbative discussion. In Section
\ref{secdual}, we discuss the temperature modification to the
perturbative and non-perturbative BPS spectra in $D-1=5$ and $D-1=4$
dimensions\footnote{As already mentioned, a $D$-dimensional
theory at finite temperature can as well be studied as a
$(D-1)$-dimensional theory, hence this notation.}. Section
\ref{seceff} presents the derivation of the effective Lagrangian for
the potentially tachyonic states, as a four-dimensional supergravity
theory, and the discussion of the minima of the scalar potential.
Section \ref{sechighphase} provides a detailed discussion of the
high-temperature phase found in perturbative heterotic strings, using
the effective supergravity theory. We show that a linear dilaton is a
background of the effective supergravity, we study the structure of
the mass spectrum, the fate of supersymmetry, and consider various
limits in this background. This phase is further discussed in Section
\ref{sechighphase2}, in the framework of non-critical strings with
${\cal N}_{sc}=2$ or ${\cal N}_{sc}=4$ superconformal symmetry. We
demonstrate the existence of massless excitations in twenty-eight
$N_4=2$ hypermultiplets. We conclude in Section \ref{secconclusions}.

\section{Perturbative analysis}\label{secpert}
\setcounter{equation}{0}

To construct the thermal partition function of a system of fields,
spin-statistics requires the boundary condition around the $S^1$
circle to be modified according to
$$
\Psi(t+2L\pi R)=(-1)^{La}\Psi(t)\, .
$$
Under a $2\pi$ rotation ($L=1$) of Euclidean time, bosons ($a=0$) are
periodic while fermions ($a=1$) are antiperiodic. The generalization
to (perturbative) string theory is dictated by modular invariance. It
replaces the above sign with \cite{AW}--\cite{AK}
$$
(-1)^{La+nb+\delta Ln}
$$
for a state with winding numbers $L$ and $n$ along the two
non-contractible loops on the world-sheet torus. Here, $a$ and $b$
denote the fermionic spin structures along these two cycles. Modular
invariance indicates that the parameter $\delta$ is equal to one for
the heterotic string and zero for the type IIA and IIB
strings\footnote{The thermal singularities of Type I strings are
the same as for Type IIB, so that we will not refer to open strings
in the sequel.}. It can be seen that the consequence of this phase is
to shift the lattice momenta of the $S^1$ string coordinate according
to the rule\footnote{Whenever $\alpha^\prime$ is not explicitly
mentioned in a formula, our convention is $\alpha^\prime=2$.}
\cite{KR, AK}
\be
\label{PLR}
P_{L,R}={1 \over R }\left[~ m+{a \over 2}-
{n\delta \over 2}\pm {nR^2 \over \alp }\right],
\ee
and to reverse the GSO projection in the odd winding number sector.

It turns out that string theories with $D$-dimensional space-time
supersymmetry look at finite temperature as if supersymmetry were
spontaneously broken in $D-1$ dimensions. Indeed, with a redefinition
of $m$, $a$ can be identified with the ($D$ dimensional) helicity
operator: $Q={\rm integer} + a/2$. Then, the states of the thermal
theory, viewed as $(D-1)$-dimensional, are mapped to those of a
supersymmetric theory compactified on $S^1$, without the temperature
spin-statistics factor $(-1)^{LQ+nb+\delta Ln}$ that induces helicity
shifts in the momenta (\ref{PLR}). Explicitly, a state of the latter
with momentum, winding and helicity charges $(m,~n,~Q)$ is mapped in
the thermal case to
\be
\label{shift}
n\rightarrow n'=n,\qquad  m \rightarrow m'= m+ {\vec e} \cdot
{\vec Q}-{\vec e}\cdot {\vec e}~ {n \over 2}, \qquad
{\vec Q}\rightarrow {\vec Q}'= {\vec Q}-{\vec e}~n\, ,
\ee
where the helicity vector $\vec Q$ is constructed in terms of the
left- and right-moving string helicities $\vec Q =(Q_L, ~Q_R)$. The
vector $\vec e=(1,0)$ in the heterotic string and $\vec e=(1,1)$ in
type II theories, and the inner product is Lorentzian: $\vec A \cdot
\vec B=A_LB_L-A_RB_R$. Note that $Q'\equiv Q_L'+Q_R'$ is the helicity
operator in $D-1$ dimensions. The above shift of charges follows from
a Lorentzian boost, which keeps invariant the combination
${\alp\over2}\,\vec P \cdot\vec P+\vec Q \cdot \vec Q $ and thus
preserves modular invariance.

The perturbative superstring mass formula can be read from the
left-movers, which carry world-sheet supersymmetry:
\beq
\label{mass0}
{1\over2}\alp {\cal M}^2 = \sum_{i=1}^4 Q_i^2 -1 +
{1\over2}\alp P_L^2 + {1\over2}\alp
{\cal M}^2_{others},
\eeq
where we have dropped the subscript $L$ for notational simplicity,
${\cal M}^2_{others}$ denotes the contributions from oscillator modes
as well as from the momenta of the remaining part of the lattice, and
$P_L$ is as in Eq.(\ref{PLR}). The four (left-) charges $Q_i$ are the
eigenvalues under the four $U(1)$ helicities acting on the
world-sheet fermions. One of them can be identified with the
contribution of left-movers to the (four-dimensional) space-time
helicity, $Q_L$, introduced in the discussion following
Eq.(\ref{shift}). These charges $Q_i$ are integers for space-time
bosons (NS states), and half-integers for space-time fermions (R
states). The supersymmetric GSO projection implies that $\sum_{i=1}^4
Q_i$ is an odd integer for NS states, while it is an even or odd
integer for R states, depending on a free choice of chirality. But in
any case, $\sum_{i=1}^4Q_i^2$ is an odd integer. The lowest BPS
states of the supersymmetric theory have ${\cal M}_{others}=0$ and
$\sum_{i=1}^4 Q_i^2=1$.

At finite temperature, the GSO projection is modified as $Q_i$ gets
shifted according to Eq.(\ref{shift}), which also affects the
momentum $P_L$. Notice that for an even winding number $n$, the
thermal modification of $P_L$ defined in Eq.(\ref{shift}) can be
regarded as a shift of $m$ and $Q$ compatible with the
(supersymmetric) GSO projection. As a consequence, the spectrum in
even $n$ sectors is not different in the thermal and supersymmetric
cases, the mass formula for the (lightest) BPS fermions, gauge bosons
and scalars with even windings $n$ remains ${\cal M}^2 = P_L^2$, with
$m$ modified as in Eq.(\ref{shift}), and tachyonic states are not
present. The situation is not the same for states with odd winding
number $n$. In this case the BPS mass formula becomes
${1\over2}\alp{\cal M}^2 = {1\over2}\alp P_L^2+n(n-2Q_L)$. {}From the
GSO condition $\sum_{i=1}^4 Q_i^2=1$, it follows that the only states
that can become tachyonic are those with $n=\pm 1$ and $Q_L=\pm
1$($=-Q_R$ for type II) \cite{AK}. They correspond to $(D-1)$-
dimensional scalars coming from the longitudinal components of the
$D$-dimensional metric.

The Hagedorn temperature is identified with the critical value of the
radius at which the first tachyonic state appears, as $2\pi R=T^{-1}$
decreases. From the above mass formula, its charges are:
\beq
\label{tachyons}
\begin{array}{rrcl}
{\rm heterotic:}&\qquad (m,n,Q) &=& \pm(-1,1,1), \crbig
{\rm type \,\, II:}& \qquad (m,n,Q_L,Q_R) &=& \pm(0,1,1,-1).
\end{array}
\eeq
The Hagedorn temperatures are $T_H={1\over\sqrt{2\alp}\pi}(\sqrt2-1)$
for the heterotic string\footnote{There is a second (higher)
critical temperature due to temperature duality,
$R\rightarrow\alp/R$, $T\rightarrow (4\pi^2\alp T)^{-1}$.} and
$T_H={1\over2\sqrt{2\alp}\pi}$ for type II theories.

The appearance of tachyons cannot take place in a perturbative
supersymmetric field theory, which behaves like the zero-winding
sector of strings; all masses (squa\-red) are increased by finite
temperature corrections, ${\cal M}^2 = P^2$, and a thermal
instability is never generated by a state becoming tachyonic at high
temperature. However, as we will see below, in non-perturbative
supersymmetric field theories such an instability can arise from
thermal dyonic modes, which behave as the odd winding string states.
Indeed, in theories with $N_4=4$ supersymmetries the BPS mass formula
is determined by the central extension of the corresponding
superalgebra \cite{N=4BPS}--\cite{KK} and dyonic field theory states
are
mapped to string winding modes \cite{Sen, KK}. Using heterotic--type
II duality, one can argue that the thermal shift of the BPS masses
modifies only the perturbative momentum charge $m$. In both heterotic
and type II perturbative strings, the thermal winding number $n$ is
not affected by the temperature shifts [see Eq.(\ref{shift})].
Since, in dimensions lower than six, heterotic--type II duality
exchanges the winding numbers $n$ of the two theories, and since the
winding number of the one theory is the magnetic charge of the other,
it is inferred that field theory magnetic numbers are not shifted at
finite temperature. This in turn indicates how to modify the BPS mass
formula at finite temperature. In Section \ref{seceff}, we will give
an independent argument based only on spontaneously broken $N_4=4$
supersymmetry and the nature of BPS states.

\section{String duality and BPS spectrum}\label{secdual}
\setcounter{equation}{0}

Our goal is to study six- and five-dimensional string theories with
$N_4=4$ supersymmetry, at finite temperature. The heterotic string is
then compactified on $T^4$ and type II theories on $K_3$. In six
dimensions, there is an S-duality that relates heterotic and type IIA
strings. Upon compactification to five dimensions on a circle, type
IIA and IIB theories are related by a T-duality.

In view of the observation that the thermal spectrum is obtained by
the modification (\ref{shift}) applied to the spectrum of the
supersymmetric theory with the temperature replaced by an ordinary
circle, we start by describing the supersymmetric BPS spectrum in
five and four dimensions. The states that can induce a thermal
instability are charged under the Kaluza-Klein $U(1)$. Their mass
depends on the temperature radius $R=(2\pi T)^{-1}$. The mass formula
from the heterotic point of view and in $\alpha_H^\prime$ units is
\beq
\label{mass1}
{\cal M}^2 =
\left( {m\over R} + {n R\over\alp_H} +
{\ell R\over\lambda_H^2 \alp_H} \right)^2,
\eeq
where $m$ and $n$ are the circle momentum and winding numbers, $\ell$
is the non-perturba\-ti\-ve wrapping number for the heterotic
five-brane around $T^4\times S^1$, with tension $\lambda_H^{-2}$ in
$\alpha^\prime_H$ units, and $\lambda_H$ is the string coupling in
six dimensions\footnote{ The six-dimensional gravitational action
in the string frame is $-{1\over2}~({\alpha_H^\prime})^{-2}\int d^6x
\,\lambda_H^{-2}eR$, so that $\lambda_H$ is dimensionless.
Traditionally, $\lambda_H$ is related to the dilaton by $\lambda_H^2
= e^{2\phi}$.}. The combination
\beq
\label{g5is}
g_5^2 = {\alpha^\prime_H\over R}\lambda_H^2
\eeq
is the five-dimensional string coupling.

Performing an S-duality in Eq.(\ref{mass1}),
\beq
\label{sdual}
\lambda_H = {1\over\lambda_{IIA}}, \qquad
\lambda_H^2 \alpha^\prime_H = \alpha^\prime_{II} ,
\eeq
we find the mass formula for type IIA strings:
\beq
\label{mass2}
{\cal M}^2 = \left( {m\over R} +
{nR\over\alpha^\prime_{II}\lambda_{IIA}^2}
+ {\ell R\over\alp_{II}} \right)^2.
\eeq
The momentum and winding numbers are now $m$ and $\ell$, while $n$ is
the wrapping number for the Neveu-Schwarz type IIA five-brane around
$K_3\times S^1$.

{}From the six-dimensional viewpoint, the first term, $m/R$, is the
Kaluza--Klein momentum, while the last two terms correspond to BPS
(dyonic) strings with tension
\beq
T_{p,q}={p\over \alp_H}+{q\over\alp_{II}}\, ,
\label{BPSstr}
\eeq
where $p,q$ are relatively primes, so that $(n,\ell)=k\,(p,q)$. The
common divisor $k$ defines the wrapping of the $T_{p,q}$ string
around $S^1$. On the type IIA side, $q$ is the charge of the
fundamental string and $p$ the magnetic charge of the solitonic
string obtained by wrapping the NS five-brane around $K_3$. These
$T_{p,q}$ strings cannot become tensionless since they are never
associated to vanishing cycles of the internal manifold.
Consequently, their tension is always positive and $p$, $q$ must be
non-negative integers. On the other hand, for $\ell=0$, the heterotic
GSO projection implies $mn\ge 0$, while for $n=0$ the type IIA
projection implies $ml\ge 0$. More generally, the $T_{p,q}$ string
implies that $mk\ge0$ \cite{DVV}.

Following the procedure described above, the temperature deformation
transforms Eq.(\ref{mass1}) into:
\beq
\label{mass3}
{\cal M}^2_T = \left( {m+Q'+{kp\over2}\over R} +
k~T_{p,q}~R \right)^2 - 2 ~T_{p,q}~ \delta_{k,\pm1} ~\delta_{Q',0}\,,
\eeq
where $Q'$ is the helicity operator in $D-1=5$ dimensions [see
Eq.(\ref{shift})]. In fact, this formula reproduces the perturbative
result for both heterotic and type IIA theories, as specified by
Eq.(\ref{shift}). In the heterotic perturbative limit
$\lambda_H\rightarrow 0$, only the $\ell=0=q$ states survive, while
in the type IIA perturbative limit $\lambda_{II}\to 0$, only the
$n=0=p$ states survive. Note that in the general case of a $T_{p,q}$
string with the temperature deformation, the condition $mk\ge0$
becomes $mk\ge-1$ because of the inversion of the GSO projection.

It follows from Eq.(\ref{mass3}) that if the heterotic coupling
$\lambda_H$ is smaller than the critical value
\beq
\label{crit}
\lambda_H^c = {\sqrt2+1\over2}
\eeq
the first tachyon has $(m,n,\ell,Q')=\pm(-1,1,0,0)$ and it appears at
$R=\sqrt{\alpha^\prime_H/2}(\sqrt2+1)$, which corresponds to the
heterotic Hagedorn temperature. On the other hand, if the heterotic
theory is strongly-coupled, $\lambda_H>\lambda_H^c$, the first
tachyon has $(m, n, \ell,Q') = \pm(0,0,1,0)$, and the critical radius
is $R=\sqrt{2\alp_H}\,\lambda_H = 2\sqrt{\alp_{II}/2}$, which
corresponds to the type IIA Hagedorn temperature. Besides the above
two would-be tachyons, mass formula (\ref{mass3}) leads in general to
two series of potentially tachyonic states with $m=-1$:
\beq
\label{Tpq1}
\begin{array}{rrrcl}
p=1,& \qquad \forall  q\,:& \qquad\qquad R &=&
\displaystyle{\left({\sqrt2\pm1 \over
\sqrt2}\right) {1\over\sqrt{T_{1,q}}}}, \crbig
p=2,& \qquad \forall  q\,\,{\rm odd}\,:& \qquad\qquad R &=&
\displaystyle{\sqrt{2\over T_{2,q}}}
\end{array}
\eeq
(which includes the first heterotic tachyon with $p=1, q=0$). The
critical temperature $(2\pi R)^{-1}$ for each of the states in both
series is always higher than the lowest Hagedorn heterotic
temperature while, as discussed above, the type IIA Hagedorn
temperature first appears when the heterotic coupling exceeds the
critical value (\ref{crit}).

In order to include type IIB strings, we need to discuss
five-dimensional theories at finite temperature, taking into account
the compactification radius $R_6$ from six to five dimensions. Type
IIA and IIB strings are then related by the inversion of $R_6$. The
extension to four dimensions of the mass formula (\ref{mass3}) is
straightforward. It depends on three parameters, the string coupling
$g_H$, the temperature radii $R$ and $R_6$. It is convenient to
introduce the three combinations
\beq
\label{stuare}
t = {RR_6\over\alpha^\prime_H},
\qquad u = {R\over R_6}, \qquad  s= g_H^{-2} = {t\over\lambda_H^2},
\eeq
in terms of which the BPS mass formula in the $N_4=4$ supersymmetric
case reads \cite{KK}:
\beq
\label{mass4}
\begin{array}{rcl}
{\cal M}^2 &=& \displaystyle{{\left|
m+ntu +i (m^\prime u + n^\prime t) + is \left[
\tilde m + \tilde n tu - i(\tilde m^\prime u +\tilde n^\prime
t)\right]
\right|^2 \over \alpha_H^\prime tu}}
\crbig
&=& \displaystyle{
\left[ {m\over R}+{nR\over\alpha_H^\prime}
+ g_H^{-2}\left({\tilde m^\prime\over R_6}+{\tilde n^\prime
R_6\over\alpha_H^\prime} \right) \right]^2
+\left[{m^\prime\over R_6}+{n^\prime R_6\over\alpha_H^\prime}
+ g_H^{-2}\left({\tilde m\over R}+{\tilde nR\over\alpha_H^\prime}
\right)\right]^2 .}
\end{array}
\eeq
In this expression, the integers $m,n,m^\prime, n^\prime$ are the
four electric momentum and winding numbers, corresponding to the four
$U(1)$ charges from $T^2$ compactification. The numbers $\tilde m,
\tilde n, \tilde m^\prime, \tilde n^\prime$ are their magnetic
non-perturbative partners, from the heterotic point of view.

The mass formula (\ref{mass4}) has been defined for heterotic
variables. To exhibit the relation with the type IIB theory, we
rewrite the above mass formula in terms of type IIA variables
(\ref{sdual}) and perform a T-duality
\beq
\label{Tdual}
R_6 = {\alpha^\prime_{II}\over R_6^B}, \qquad
\lambda_{IIA} = \lambda_{IIB} {\sqrt{\alpha^\prime_{II}}\over R_6^B}.
\eeq
However, since\footnote{$\kappa$ is the {\it four-dimensional}
gravitational coupling, $\kappa=\sqrt{8\pi}M_P^{-1}= (2.4\times
10^{18}\,\,{\rm GeV})^{-1}$.}
$$
R^2 = \alpha^\prime_H tu  = 2\kappa^2 stu
$$
and $R$ is by construction identical in all three string theories,
the mass formula (\ref{mass4}) is invariant under the exchanges
$s\leftrightarrow t$, $s\leftrightarrow u$ and $t\leftrightarrow u$.
These operations correspond respectively to heterotic--IIA, IIA--IIB
and heterotic--IIB dualities. The mass formula will then apply to all
three theories, provided $s$, $t$ and $u$ are defined as in
Eqs.(\ref{stuare}), but in terms of the appropriate variables
$\alpha^\prime$, $R_6$ and $\lambda$ in each theory.

The five-dimensional IIA formula (\ref{mass1}) is reobtained by
choosing first $m^\prime = n^\prime= 0$, which removes the second
torus, and then taking the limit $R_6\to\infty$ with $\lambda_H$ kept
fixed. This limit implies $\tilde m =\tilde n = \tilde n^\prime =0$,
and $\tilde m^\prime$ is identified with $\ell$ in Eq.(\ref{mass1}).

The five-dimensional IIB theory is obtained by taking $\tilde m =
\tilde n = 0$ and the limit $R_6^B\to \infty$, which implies $n =
m^\prime = n^\prime = 0$. We reobtain Eq.(\ref{mass2}) with IIB
variables, $\ell=\tilde n^\prime$, $n=\tilde m^\prime$ and $m$
unchanged. Similarly, the finite temperature mass formula is
identical to Eq.(\ref{mass3}) with the same identification.

Finally the four-dimensional thermal mass formula is obtained from
Eq.(\ref{mass4}) by replacing $m$ by $m+Q'+n/2$:
\beq
\label{mass5}
{\cal M}^2_T =
\left({m+Q'+{kp\over 2}\over R}+
k~T_{p,q,r}~R\right)^2-2 ~T_{p,q,r}~\delta_{|k|,1}
{}~\delta_{Q',0}\, ,
\eeq
where we have set $m'=n'={\tilde m}={\tilde n}=0$ corresponding to
the lightest states, and we defined $k$ as before, as the common
divisor of $(n,{\tilde m^\prime},{\tilde n^\prime})\equiv k(p,q,r)$.
Then $T_{p,q,r}$ is an effective string tension
$$
T_{p,q,r}={p\over\alpha_H^\prime}
+{q\over\lambda_H^2\alpha_H^\prime}
+{r R_6^2\over\lambda_H^2(\alpha_H^\prime)^2}\, .
$$
Note that $\tilde m^\prime=kq$ corresponds to the wrapping number of
the heterotic five-brane around $T^4\times S^1_R$ as in five
dimensions, while $\tilde n^\prime=kr$ corresponds to the same
wrapping number after performing a T-duality along the $S^1_{R_6}$
direction, which is orthogonal to the five-brane. As we discussed in
the previous section, all winding numbers $n, {\tilde m^\prime},
{\tilde n^\prime}$ correspond to magnetic charges from the field
theory point of view. Their masses are proportional to the
temperature radius $R$ and are not thermally shifted.

A nicer expression of the effective string tension $T_{p,q,r}$ is:
\beq
\label{Tpqr}
T_{p,q,r}={p\over\alpha_H^\prime}
+{q\over\alpha_{IIA}^\prime}
+{r\over\alpha_{IIB}^\prime}\,,
\eeq
where the various $\alpha^\prime$ are
\beq
\label{alphaare}
\alpha_H^\prime = 2\kappa^2 s, \qquad\qquad
\alpha_{IIA}^\prime = 2\kappa^2 t, \qquad\qquad
\alpha_{IIB}^\prime = 2\kappa^2 u, \qquad\qquad
\eeq
when expressed in Planck units. Note that $\alpha_{IIB}^\prime$
defines a new type II theory obtained by heterotic T-duality with
respect to $R_6$, in contrast to the type IIA--IIB T-duality
(\ref{Tdual}). In the following we will refer to this new theory as
perturbative IIB.

We stress here that $p,q,r$ are all non-negative relatively prime
integers. This follows from the constraints $n{\tilde m^\prime}\ge
0$, $n{\tilde n^\prime}\ge 0$ and ${\tilde m^\prime}{\tilde
n^\prime}\ge 0$, which are a consequence of the BPS conditions and
the $s\leftrightarrow t\leftrightarrow u$ duality symmetry in the
undeformed supersymmetric theory. Futhermore, $mk\ge -1$ because of
the inversion of the GSO projection in the temperature-deformed
theory. Using these constraints, it is straightforward to show that
in general there are two potential tachyonic series with $m=-1$ and
$p=1,2$, generalizing the five-dimensional result (\ref{Tpq1}):
\beq
\label{Tpqr2}
\begin{array}{rrrcl}
p=1,& \qquad \forall  (q,r) ~ {\rm relat. ~primes}:& \qquad\qquad R
&=&
\displaystyle{\left({\sqrt2\pm1 \over
\sqrt2}\right) {1\over\sqrt{T_{1,q,r}}}}, \crbig
p=2,& \qquad \forall  (p,q,r) ~{\rm relat. ~primes}:& \qquad\qquad R
&=&
\displaystyle{\sqrt{2\over T_{2,q,r}}}
\end{array}
\eeq

One of the perturbative heterotic type IIA, or type IIB potential
tachyons correspond to a critical temperature that is always lower
than the above two series. The perturbative Hagedorn temperatures
are:
$$
\begin{array}{lll}
{\rm heterotic \,\,  tachyon:}\qquad &
(m,n,Q')=\pm(-1,1,0), \qquad
&2\pi T = \left({\sqrt2-1}\right){\sqrt{2\over\alpha^\prime_H}};
\crbig
{\rm type \,\, IIA \,\, tachyon:}\qquad &
(m,\tilde m^\prime,Q')=\pm(0,1,0), \qquad
&2\pi T = {1\over\sqrt{2\alpha^\prime_{IIA}}};
\crbig
{\rm type \,\, IIB \,\, tachyon:}\qquad &
(m,\tilde n^\prime,Q')=\pm(0,1,0) , \qquad
&2\pi T = {1\over\sqrt{2\alpha^\prime_{IIB}}}.
\end{array}
$$

This discussion shows that the temperature modification of the mass
formula inferred from perturbative strings and applied to the
non-perturbative BPS mass formula produces the appropriate
instabilities in terms of Hagedorn temperature. We will now proceed
to show that it is possible to go beyond the simple enumeration of
Hagedorn temperatures. We will construct an effective supergravity
Lagrangian that allows a study of the nature of the non-perturbative
instabilities and the dynamics of the various thermal phases.

\section{Four-dimensional effective supergravity}\label{seceff}
\setcounter{equation}{0}

In the previous section, we have studied, at the level of the mass
formula for $N_4=4$ BPS states, the appearance of tachyonic states
generating thermal instabilities. To obtain information on dynamical
aspects of these instabilities, we now construct the full
temperature-dependent effective potential associated with the
would-be tachyonic states.

Our procedure to construct the effective theory is as follows. We
consider five-dimensional $N_4=4$ theories at finite temperature.
They can then effectively be described by four-dimensional theories,
in which supersymmetry is spontaneously broken by thermal effects.
Since we want to limit ourselves to the description of instabilities,
it is sufficient to only retain, in the full $N_4=4$ spectrum, the
potentially massless and tachyonic states. This restriction will lead
us to consider only spin 0 and 1/2 states, the graviton and the
gravitino\footnote{The four gravitinos remain degenerate at
finite temperature; it is then sufficient to retain only one of
them.}. This sub-spectrum is described by an $N_4=1$ supergravity
with chiral multiplets\footnote{When considering six-dimensional
theories at finite temperature, one is similarly led to consider an
$N_4=2$ theory with vector and hypermultiplets. We will briefly
return to this point later in Section 4.3.}.

The scalar manifold of a generic, {\it unbroken}, $N_4=4$ theory is
\cite{DF}--\cite{FK}
\beq
\label{manif1}
\left({Sl(2,R) \over U(1)}\right)_S \times \, G/H, \qquad\qquad
G/H =
\left({SO(6,r+n)\over SO(6)\times SO(r+n)}\right)_{T_I,\phi_A}.
\eeq
The manifold $G/H$ of the $N_4=4$ vector multiplets
naturally splits into a part that includes the $6r$
moduli $T_I$, and a second part which includes the infinite number
$n\rightarrow\infty$ of BPS states $\phi_A$.

In the manifold $G/H$, we are only interested in keeping the six BPS
states $Z_A^\pm$, $A=1,2,3$, which, according to our discussion in
the previous section, generate thermal instabilities in heterotic,
IIA and IIB strings. For consistency, these states must be
supplemented by two moduli $T$ and $U$ among the $T_I$'s. We consider
heterotic and type II strings respectively on $T^4\times S^1_6\times
S^1_5$ and $K_3\times S^1_6\times S^1_5$, where $S^1_6$ is a trivial
circle and $S_5^1$ is the temperature circle. The moduli $T$ and $U$
describe the $T^2\equiv S^1_5\times S^2_6$ torus. Thus, $r+n = 8$ in
the $N_4=4$ manifold (\ref{manif1}). To construct the appropriate
truncation of the scalar manifold $G/H$, which only retains the
desired states of $N_4=1$ chiral multiplets, we use a $Z_2\times Z_2$
subgroup contained in the $SO(6)$ R-symmetry of the coset $G/H$. This
symmetry can be used as the point group of an $N_4=1$ orbifold
compactification, but we will only use it for projecting out
non-invariant states of the $N_4=4$ theory\footnote{Only
untwisted states would contribute to thermal instabilities.} with
$r+n=8$.

A single $Z_2$ would split $H=SO(6)\times SO(8)$ in $[SO(2)\times
SO(2)]\times[SO(4)\times SO(6)]$, and the scalar manifold would
become
\beq
\label{manif2}
\begin{array}{l}
\hspace{1.cm}\displaystyle{\left({Sl(2,R) \over U(1)}\right)_S \times
\left({SO(2,2)\over SO(2)\times SO(2)}\right)_{TU} \times
\left({SO(4,6)\over SO(4)\times SO(6)}\right)_{\phi_A}} \crbig
\hspace{.6cm}\displaystyle{
=\left({Sl(2,R) \over U(1)}\right)_S \times
\left({Sl(2,R) \over U(1)}\right)_T \times
\left({Sl(2,R) \over U(1)}\right)_U \times
\left({SO(4,6)\over SO(4)\times SO(6)}\right)_{\phi_A},}
\end{array}
\eeq
At this stage, the theory would have $N_4=2$ supersymmetry and the
first three factors in the scalar manifold are vector multiplet
couplings with prepotential ${\cal F}=iSTU/X_0$. The last one is a
quaternionic coupling of hypermultiplets. The second $Z_2$ projection
acts on this factor and reduces it to
\beq
\label{manif3}
\left({SO(2,3)\over SO(2)\times SO(3)}\right)_{Z_A^+}\times
\left({SO(2,3)\over SO(2)\times SO(3)}\right)_{Z_A^-},
\qquad A=1,2,3.
\eeq
This is a K\"ahler manifold for chiral multiplets coupled to $N_4=1$
supergravity \cite{CFGVP}. The second $Z_2$ projection also truncates
$N_4=2$ vector multiplets into $N_4=1$ chiral multiplets.

The structure of the truncated scalar manifold indicates that the
K\"ahler potential can be written as
\beq
\label{Kis}
\begin{array}{rcl}
K &=& -\log(S+S^*)-\log(T+T^*)-\log(U+U^*) \crbig
&&-\log Y(Z_A^+,Z_A^{+*}) -\log Y(Z_A^-,Z_A^{-*}),
\end{array}
\eeq
with
\beq
\label{Yis}
Y(Z_A^\pm,Z_A^{\pm*}) = 1 -2Z_A^\pm Z_A^{\pm*} + (Z_A^\pm
Z_A^\pm)(Z_B^{\pm*}Z_B^{\pm*}).
\eeq
This choice is a solution to the $N_4=4$ constraints. For the
$S$-manifold $SU(1,1)/U(1)$ $\sim Sl(2,R)/U(1)$, the constraint is
\beq
\label{Sconst}
|\varphi_0|- |\varphi_1|^2 =1/2.
\eeq
The solution we use reads
\beq
\label{Sconstsol}
\varphi_0 - \varphi_1= {1\over(S+\ov S)^{1/2}}, \qquad\qquad
\varphi_0 + \varphi_1= {S\over(S+\ov S)^{1/2}}.
\eeq
For an $SO(2,n_I)/SO(2)\times SO(n_I)$ manifold, the constraints are
\beq
\label{2nconst}
\begin{array}{rcl}
|\sigma_I^1|^2 + |\sigma_I^2|^2 - |\vec\phi_I|^2 &=& 1/2, \crbig
(\sigma_I^1)^2 + (\sigma_I^2)^2 - (\vec\phi_I)^2 &=& 0,
\end{array}
\eeq
where $\vec\phi_I$ has $n_I$ components and we introduced the index
$I$, with values $0,+,-$, because we have three such manifolds, $I=0,
n_I=2$ for the moduli $T$ and $U$, $I=\pm, n_I=3$ for the winding
states
$Z_A^+$ and $Z_A^-$ ($A=1,2,3=n_\pm$).
The standard parametrization for the $TU$ manifold, analogous to
the choice of $S$, corresponds to the solution
\beq
\label{TUsol}
\sigma_0^1 = {1+TU\over 2Y_{TU}^{1/2}}, \qquad
\sigma_0^2 = i{T+U\over 2Y_{TU}^{1/2}}, \qquad
\phi_0^1 = {1-TU\over 2Y_{TU}^{1/2}}, \qquad
\phi_0^2 = i{T-U\over 2Y_{TU}^{1/2}}, \qquad
\eeq
with
$$
Y_{TU} = (T+\ov T)(U+\ov U).
$$
For the $I=\pm$ manifolds $SO(2,3)/SO(2)\times SO(3)$, a convenient
parametrization is
\beq
\label{Zsol}
\sigma_\pm^1 = {1+ (Z_A^\pm)^2 \over 2 Y_\pm^{1/2}}, \qquad
\sigma_\pm^2 = i{1- (Z_A^\pm)^2 \over 2 Y_\pm^{1/2}}, \qquad
\phi_\pm^A = {Z_A^\pm \over Y_\pm}, \qquad A=1,2,3,
\eeq
where $Y_\pm$ has been defined in Eq.(\ref{Yis}). The K\"ahler
function, which defines the $N_4=1$ supergravity theory, can be
determined by directly comparing the gravitino mass terms in the
$N_4=1$ Lagrangian with the similar term obtained after $Z_2\times
Z_2$ truncation of the $N_4=4$ theory:
\beq
\label{gravmass}
e^{K/2}W = (\varphi_0-\varphi_1)f_{ijk}\Phi_0^i\Phi_+^j\Phi_-^k
+ (\varphi_0+\varphi_1)\tilde f_{ijk}\Phi_0^i\Phi_+^j\Phi_-^k,
\eeq
where
$$
\Phi_{0,\pm}^i = (\,\sigma^1_{0,\pm}\, , \,\sigma^2_{0,\pm}\,
, \,\vec \phi_{0,\pm}\,).
$$
The structure constants $f_{ijk}$ and $\tilde f^{ijk}$ characterize
the self-couplings of the $N_4=4$ vector multiplets. In this sense,
they define the {\it gauging} of the $N_4=4$ theory
\cite{dR,GP,susybreak,PZ}. They induce a scalar potential that can,
when appropriately chosen, spontaneously break supersymmetry
\cite{AK}.

The solutions (\ref{Sconstsol}), (\ref{TUsol}) and (\ref{Zsol}) to
the $N_4=4$ constraints indicate that the non-analytic contribution
to the gravitino mass (\ref{gravmass}) is $[(S+\ov S)
Y_{TU}Y_+Y_-]^{-1/2}$. This is identified with $e^{K/2}$ and leads to
the expression (\ref{Kis}) of the $N_4=1$ K\"ahler potential. The
analytic superpotential then is
$$
W = [(S+\ov S)Y_{TU}Y_+Y_-]^{1/2} \left[(\varphi_0-\varphi_1)
f_{ijk}\Phi_0^i\Phi_+^j\Phi_-^k
+ (\varphi_0+\varphi_1)\tilde f_{ijk}\Phi_0^i\Phi_+^j\Phi_-^k\right],
$$
using the solutions to the $N_4=4$ constraints, once the gauging has
been specified.

The superpotential for generic $N_4=4$ strings, after the $Z_2\times
Z_2$ truncation to $N_4=1$, is:
\beq
\label{susyW}
\begin{array}{rcl}
W_{susy}&=&
\left[ m_1(\sigma_0^1+\phi_0^1) + n_1(\sigma_0^1-\phi_0^1)\right]
(\varphi_0-\varphi_1) \phi_+^{(m_1,n_1)} \phi_-^{(m_1,n_1)} \crbig
&&
+\left[ m_2(\sigma_0^2+\phi_0^2) + n_2(\sigma_0^2-\phi_0^2)\right]
(\varphi_0-\varphi_1) \phi_+^{(m_2,n_2)} \phi_-^{(m_2,n_2)} \crbig
&&
+\left[ \tilde m_1(\sigma_0^1+\phi_0^1)+\tilde
n_1(\sigma_0^1-\phi_0^1)\right]
(\varphi_0+\varphi_1) \tilde\phi_+^{(\tilde m_1,\tilde n_1)}
\tilde\phi_-^{(\tilde m_1,\tilde n_1)} \crbig
&&
+\left[ \tilde m_2(\sigma_0^2+\phi_0^2)+\tilde
n_2(\sigma_0^1-\phi_0^1)\right]
(\varphi_0+\varphi_1) \tilde\phi_+^{(\tilde m_2,\tilde n_2)}
\tilde\phi_-^{(\tilde m_2,\tilde n_2)} .
\end{array}
\eeq
The contributions proportional to $\varphi_0-\varphi_1 = (S+\ov
S)^{-1/2}$ give rise to the perturbative $T^2$ heterotic string
spectrum, provided the numerical coefficients $m_1,m_2,n_1,n_2$ are
equal to the momentum and winding charges \cite{GP,AK}. These
contributions define the structure constants $f_{ijk}$ in
Eq.(\ref{gravmass}). The contributions proportional to
$\varphi_0+\varphi_1 = S(S+\ov S)^{-1/2}$ provide the mass spectrum
of the non-perturbative magnetic $T^2$ torus, and correspond to the
structure constants $\tilde f_{ijk}$. This superpotential summarizes
the complete BPS mass spectrum valid for all (truncated) $N_4=4$
strings (heterotic, type IIA and type IIB). It is worth recalling
that the expression of the superpotential, together with the K\"ahler
potential $K$, defines not only mass terms, but the full scalar
sector and its coupling to $N_4=1$ supergravity. This allows to
examine in principle the vacuum structure far from large (or small)
values of $S+\ov S$.

The superpotential (\ref{susyW}) does not, however, break
supersymmetry and is not appropriate to a finite-temperature theory.
In general, breaking $N_4=4$ supergravity requires a gauging with
non-zero structure constants cubic in the compensating fields
$\sigma^{1,2}_{0+-}$. The appropriate finite-temperature gauging can
be found either in field theory from the $N_4=4$ thermal spectrum or
by examining the heterotic string spectrum at finite temperature,
which corresponds to the Scherk--Schwarz gauging
\cite{susybreak,PZ,AK}. The result is to add
\beq
\label{Wbr}
\delta W =
e(\varphi_0-\varphi_1)(\sigma_0^1+\phi_0^1)\sigma^1_+\sigma_-^1
\eeq
to $W_{susy}$. The numerical coefficient $e$ is fixed by the thermal
mass of the gravitino. In addition, the coefficient
$m_1$ is shifted at finite
temperature, according to the rule (\ref{shift}) discussed in
Section \ref{secpert}. Inserting the representation of the scalar
field, and truncating the spectrum to retain only the
odd winding states, we find the superpotential
\beq
\label{Wis}
\begin{array}{rcl}
W &=& 2\sqrt2 \biggl[ {1\over2}(1-Z_A^+Z_A^+)(1-Z_B^-Z_B^-)
 \crbig
&& +\,(TU-1)Z_1^+Z_1^- + SUZ_2^+Z_2^- +
STZ_3^+Z_3^-  \biggr].
\end{array}
\eeq

As a check, the same result can be derived at the $N_4=2$ level,
considering a single $Z_2$ truncation of the $N_4=4$ theory. The
vector multiplets ate $S$, $T$ and $U$ with manifold
$[Sl(2,R)/U(1)]^3$ with prepotential \cite{dWLVP, C7}
\beq
\label{Fis}
{\cal F}(S,T,U) = i{STU\over X_0},
\eeq
where $X_0$ is the compensating scalar in the (superconformal)
vector multiplet describing the $N_4=2$ graviphoton. The
superpotential in $N_4=1$ language has the general form
\cite{FKLZ}
$$
W = \gamma (m_IX^I-n^I{\cal F}_I) \Phi^i_+\Phi^i_-, \qquad\qquad
{\cal F}_I= {\partial\over\partial X^I}{\cal F},
$$
with a numerical constant $\gamma$. The index $I$ runs over $X^0$,
$S$, $T$ and $U$. After performing the algebra, we take the
Poincar\'e gauge $X^0=1$. The part of the superpotential that leaves
supersymmetry unbroken corresponds to the $\phi_{+-}^A$ terms in
$\Phi^i_+\Phi^i_-$. They provide the entire BPS mass terms, electric
and magnetic. The compensator contributions $\sigma_+^i\sigma_-^i$
($i=1,2$) provide the desired breaking terms and correspond to
$\delta W$, Eq.(\ref{Wbr}). The $N_4=2$ formulation can be useful to
examine the finite temperature non-perturbative behaviour of theories
based upon more general prepotentials than (\ref{Fis}), such as $K3$
compactifications of heterotic strings or Calabi--Yau threefolds of
type II strings.

\subsection{The scalar potential}

We now analyse the thermal effective potential and its instabilities.
It follows from the general expression of $N_4=1$ supergravity
coupled to chiral multiplets, for a given K\"ahler potential $K$ and
superpotential $W$.

Positivity of the kinetic energies and the form of the K\"ahler
potential (\ref{Kis}) impose that $s,t,u >0$, as well as non-trivial
conditions on $Z_A^\pm$. In particular,
$$
\sum_A (\Re\, Z_A^\pm)^2 < 1.
$$

The scalar potential\footnote{Using the standard notation $K_i =
{\partial K\over\partial z^i}$, \dots Scalar fields are
dimensionless.}
$$
V = \kappa^{-4}
e^K\left[ (K^{-1})^i_j(W_i+WK_i)(\ov W^j + \ov WK^j) -3|W|^2\right],
$$
is of course complicated. It can, however, be written in a closed
form, which is given in the Appendix. From the analysis of the mass
matrices around the vacuum $Z_A^\pm=0$, it is apparent that the
discussion of thermal instabilities and of possible phase transitions
only relies upon the scalar field directions
\beq
\label{stu}
s=\Re S, \qquad t=\Re T, \qquad u =\Re U,
\eeq
and, in the winding modes sector,
$$
\Re Z_A^+ = \Re Z_A^- \equiv z_A.
$$
Important simplifications in the potential occur then. For instance,
the winding mode K\"ahler metric becomes diagonal:
$$
(K^\pm)^A_B = {\partial^2 K\over\partial Z_A^\pm \partial \overline
Z_B^\pm}
={2 \over(1-x^2)^2}\,\delta^B_A, \qquad\qquad x^2 = \sum_A z_A^2,
$$
and the kinetic terms of the scalars $z_A$ are
\beq
\label{zAkin}
{4\over(1-x^2)^2}(\partial_\mu z_A)(\partial^\mu z_A).
\eeq
It is interesting to observe that the resulting scalar potential is a
simple fourth-order polynomial when expressed in terms of new field
variables $H_A$, taking values on the entire real axis,
\beq
\label{newdef}
H_A = {z_A\over 1-x^2}, \qquad\qquad A=1,2,3.
\eeq
Defining also
\beq
\label{xidef}
\xi_1 = tu, \qquad \xi_2 = su, \qquad \xi_3 = st
\eeq
($\xi_i>0$), the potential can be nicely rewritten as
\beq
\label{pot4}
\begin{array}{rcl}
V &=& V_1 + V_2 + V_3, \crbig
\kappa^4 V_1 &=& \displaystyle{4\over s}\left[
(\xi_1+\xi_1^{-1})H_1^4
+{1\over4}(\xi_1-6+\xi_1^{-1})H_1^2 \right],
\crbig
\kappa^4 V_2 &=& \displaystyle{4\over t}
\left[ \xi_2H_2^4 +{1\over4}(\xi_2-4)H_2^2\right],
\crbig
\kappa^4 V_3 &=& \displaystyle{4\over u}
\left[ \xi_3H_3^4 +{1\over4}(\xi_3-4)H_3^2\right].
\end{array}
\eeq
This expression displays the duality properties
\begin{itemize}
\item[] $\xi_1 \,\,\rightarrow\,\, \xi_1^{-1}$:
heterotic temperature duality;
\item[] $t \,\,\leftrightarrow\,\, u$, $H_2 \,\,\leftrightarrow\,\,
H_3$:
IIA--IIB duality.
\end{itemize}

Since at $H_i=0$, the K\"ahler metric is $4\delta^A_B$, the scalar
potential is normalized according to $V = {4\kappa^2}\sum_A m_A^2
H_A^2 + \ldots$ The masses $m_A^2$ correspond to the mass formula for
the heterotic, IIA and IIB tachyons respectively:
$$
\begin{array}{rcccll}
m_1^2 &=& \displaystyle{{1\over4\kappa^2
s}\left[\xi_1^{-1}+\xi_1-6\right]}
&=&\displaystyle{
{1\over2\alpha^\prime_H}\left[ {\alpha^\prime_H\over 2R^2}
+{2R^2\over\alpha^\prime_H} -6\right]} &\qquad\qquad{\rm (heterotic)
},
\crbig
m_2^2 &=& \displaystyle{{1\over4\kappa^2 t}\left[ \xi_2-4 \right]}
&=& \displaystyle{{1\over2\alpha^\prime_{IIA}}
\left[ {2R^2\over\alpha^\prime_{IIA}}
-4\right]}&\qquad\qquad{\rm (IIA)},
\crbig
m_3^2 &=& \displaystyle{{1\over4\kappa^2 u}\left[ \xi_3-4 \right]}
&=& \displaystyle{{1\over2\alpha^\prime_{IIB}}
\left[ {2R^2\over\alpha^\prime_{IIB}}
-4\right]} &\qquad\qquad{\rm (IIB)}.
\end{array}
$$
The three $\alpha^\prime$ scales are clearly as in
Eqs.(\ref{alphaare}). Notice that the variables $s$, $t$ and $u$ we
are using in the finite-temperature case are, with respect to the
case of unbroken supersymmetry, scaled by a factor $\sqrt2$ when
expressed in terms of stringy quantities:
\beq\label{xisare}
\xi_1 = {2R^2\over\alpha^\prime_H}, \qquad
\xi_2={2R^2\over\alpha^\prime_{IIA}}, \qquad
\xi_3={2R^2\over\alpha^\prime_{IIB}}.
\eeq

\subsection{Phase structure of the thermal effective
theory}\label{secphase}

The scalar potential (\ref{pot4}) derived from our effective
supergravity possesses four different phases corresponding to
specific regions of the $s$, $t$ and $u$ moduli space. Their
boundaries are defined by critical values of the moduli $s$, $t$, and
$u$ (or of $\xi_i$, $i=1,2,3$), or equivalently by critical values of
the temperature, the (four-dimensional) string coupling and the
compactification radius $R_6$. These four phases are:
\begin{enumerate}
\item The {\sl low-temperature} phase: \\
$T<(\sqrt2-1)^{1/2}/(4\pi\kappa)\,$;
\item The {\sl high-temperature heterotic} phase: \\
$T>(\sqrt2-1)^{1/2}/(4\pi\kappa)\,\,$ and $\,\,g_H^2<(2+\sqrt2)/4\,$;
\item The {\sl high-temperature type IIA} phase: \\
$T>(\sqrt2-1)^{1/2}/(4\pi\kappa)$\,\,, $\,\,g_H^2>(2+\sqrt2)/4\,\,$
and $\,\,R_6>\sqrt{\alpha^\prime_H}\,$;
\item The {\sl high-temperature type IIB} phase: \\
$T>(\sqrt2-1)^{1/2}/(4\pi\kappa)\,\,$, $\,\,g_H^2>(2+\sqrt2)/4\,\,$
and $\,\,R_6<\sqrt{\alpha^\prime_H}\,$.
\end{enumerate}
The distinction between phases 3 and 4 is, however, somewhat
academic,
since there is no phase boundary at $R_6=\sqrt{\alpha^\prime_H}$.

\subsubsection{Low-temperature phase}

This phase, which is {\it common} to all three strings,
is characterized by
\beq
\label{ltp1}
H_1=H_2=H_3=0, \qquad V_1=V_2=V_3=0.
\eeq The potential vanishes for all values of the moduli $s$, $t$ and
$u$, which are then restricted only by the stability of the phase,
namely the absence of tachyons in the mass spectrum of the scalars
$H_i$. This mass spectrum is analysed in detail in the appendix. This
leads to:
\beq
\label{ltp2}
\xi_1 > \xi_H = (\sqrt2+1)^2, \qquad
\xi_2 > \xi_A = 4, \qquad
\xi_3 > \xi_B = 4.
\eeq
{}From the above condition, it follows in particular that the
temperature must verify
\beq
\label{Tis}
T = {1\over 2\pi\kappa} \left({1\over\xi_1\xi_2\xi_3}\right)^{1/4}
< {(\sqrt2-1)^{1/2}\over4\pi\kappa}.
\eeq
Since the (four-dimensional) string couplings are
$$
s = \sqrt2 g_H^{-2}, \qquad t = \sqrt 2 g_A^{-2}, \qquad
u = \sqrt2 g_B^{-2},
$$
this phase exists in the perturbative regime of all three strings.
The relevant light thermal states are just the massless modes of the
five-dimensional $N_4=4$ supergravity, with thermal mass scaling like
$1/R \sim T$.

\subsubsection{High-temperature heterotic phase}

This phase is defined by
\beq
\label{hhp1}
\xi_H > \xi_1 > {1\over\xi_H},
\qquad\qquad \xi_2>4, \qquad\qquad \xi_3>4,
\eeq
with $\xi_H =(\sqrt2+1)^2$, as in Eq.(\ref{ltp2}). The inequalities
on $\xi_2$ and $\xi_3$ eliminates type II instabilities. In this
region of the moduli, and after minimization with respect to $H_1$,
$H_2$ and $H_3$, the potential becomes
$$
\kappa^4 V = -{1\over s}{(\xi_1+\xi_1^{-1}-6)^2\over
16(\xi_1+\xi_1^{-1})}.
$$
It has a stable minimum for fixed $s$ (for fixed $\alpha^\prime_H$)
at the minimum of the self-dual\footnote{With respect to heterotic
temperature duality.} quantity $\xi_1+\xi_1^{-1}$:
\beq
\label{highhetmin}
\xi_1=1 , \qquad H_1={1\over2} , \qquad H_2=H_3=0, \qquad
\kappa^4 V= -{1\over 2s}.
\eeq
The transition from the low-temperature vacuum is due to a
condensation of the heterotic thermal winding mode $H_1$, or
equivalently by a condensation of type IIA NS five-brane in the type
IIA picture.

At the level of the potential only, this phase exhibits a runaway
behaviour in $s$. We will show in the next section that a stable
solution to the effective action exists with non-trivial metric
and/or dilaton.

In heterotic language, $s$, $t$ and $u$ are particular combinations
of the four-dimensional gauge coupling $g_H$, the temperature $T=
(2\pi R)^{-1}$ and the compactification radius from six to five
dimensions $R_6$. The relations are
\beq
\label{hetstuare}
\begin{array}{c} \displaystyle{
s= \sqrt2g_H^{-2}, \qquad t = \sqrt2{RR_6\over\alpha^\prime_H},
\qquad u = \sqrt2{R\over R_6},}
\crbig \displaystyle{
\xi_1 = tu = {2R^2\over\alpha^\prime_H}, \qquad
\xi_2 = {2R\over g_H^2R_6}, \qquad
\xi_3 = {2RR_6\over\alpha^\prime_H g_H^2}. }
\end{array}
\eeq
As expected, $\xi_2$ and $\xi_3$ are related by radius inversion,
$R_6 \,\rightarrow\, \alpha^\prime_H R_6^{-1}$.
Then, in Planck units,
\beq
\label{hetTRuare}
R = {1\over 2\pi T} = \kappa \sqrt{stu} =
\kappa[\xi_1\xi_2\xi_3]^{1/4},
\qquad R_6 = \kappa \left({2st\over u}\right)^{1/2}
= {\sqrt2 \kappa\xi_3\over[\xi_1\xi_2\xi_3]^{1/4}}.
\eeq
The first equation indicates that the temperature, when expressed in
units of the {\it four-dimensional} gravitational coupling constant
$\kappa$ is invariant under string--string dualities.

In terms of heterotic variables, the critical temperatures
(\ref{hhp1}) separating the heterotic phases are
\beq
\label{Thetcrit}
\begin{array}{rcrrcl}
\xi_1 &=& \xi_H: \qquad&
2\pi T^<_H &=& \displaystyle{g_H\over 2^{1/4}\kappa}(\sqrt2-1),
\crbig
\xi_1 &=& {1\over\xi_H}:\qquad &
2\pi T^>_H &=& \displaystyle{g_H\over 2^{1/4}\kappa}(\sqrt2+1).
\end{array}
\eeq
In addition, heterotic phases are separated from type II
instabilities by the following critical temperatures:
\beq
\label{typeIIcrit}
\begin{array}{rrclrcl}
{\rm IIA:}\qquad& \xi_2&=&4,\qquad&2\pi T_{A} &=&
\displaystyle{R_6\over 4\sqrt2\kappa^2},
\crbig
{\rm IIB:}\qquad& \xi_3&=&4,\qquad&2\pi T_{B} &=&
\displaystyle{1\over 2 g_H^2 R_6}.
\end{array}
\eeq
Then the domain of the moduli space that avoids type II
instabilities is defined by the inequalities $\xi_{2,3}>4$.
In heterotic variables,
\beq
\label{Tineq}
2\pi T < {1\over 2\alpha^\prime_H g_H^2}\, {\rm min}\left(
R_6\,\,;\,\,
\alpha^\prime_H/R_6\right)
= {1\over 4\sqrt2\kappa^2}\, {\rm min}\left( R_6\,\,;\,\,
\alpha^\prime_H/R_6\right).
\eeq
Type II instabilities are unavoidable when $T>T_{\rm self-dual}$,
with
$$
2\pi T_{\rm self-dual} = {1\over 2g_H^2\sqrt{\alpha^\prime_H}}
= {2^{1/4}\over 4\kappa g_H}.
$$
The high-temperature heterotic phase cannot be reached\footnote{From
low heterotic temperature.} for any value of the radius $R_6$ if
$$
T^<_H > T_{\rm self-dual},
$$
or
\beq
\label{glim}
g_H^2 > {\sqrt2+1\over2\sqrt2} \,\sim \,0.8536.
\eeq
In this case, $T^<_H$ always exceeds $T_{A}$ and $T_{B}$. Only type
II thermal instabilities exist in this strong-coupling regime and the
value of $R_6/\sqrt{\alpha^\prime_H}$ decides whether the type IIA or
IIB instability will have the lowest critical temperature, following
Eq.(\ref{typeIIcrit}).

If on the other hand the heterotic string is weakly coupled,
\beq
\label{glim1}
g_H^2 < {\sqrt2+1\over2\sqrt2},
\eeq
the high-temperature heterotic phase is reached for values of the
radius $R_6$ verifying $T^<_H < T_{A}$ and $T^<_H < T_{B}$, or
\beq
\label{R6ineq}
2\sqrt2g_H^2(\sqrt2-1) < {R_6\over\sqrt{\alpha^\prime_H}} <
{1\over2\sqrt2g_H^2(\sqrt2-1)}.
\eeq
The large and small $R_6$ limits, with fixed coupling $g_H$, again
lead to either type IIA or type IIB instability.

\subsubsection{High-temperature type IIA and IIB phases}

These phases are defined by inequalities:
\beq
\label{IIphases1}
\xi_2<4 \qquad {\rm and/or}\qquad \xi_3<4.
\eeq
In this region of the parameter space, either
$H_2$ or $H_3$ become tachyonic and acquire a vacuum value:
\beq
\label{IIAphase}
H_2^2 = {4-\xi_2\over 8\xi_2}, \qquad\qquad
\kappa^4 V_2=-{1\over t}{(4-\xi_2)^2\over16\xi_2},
\eeq
and/or
\beq
\label{IIBphase}
H_3^2 = {4-\xi_3\over 8\xi_3}, \qquad\qquad
\kappa^4 V_3=-{1\over u}{(4-\xi_3)^2\over16\xi_3}.
\eeq
In contrast with the high-temperature heterotic phase, the potential
does not possess stationary values of $\xi_2$ and/or $\xi_3$, besides
the critical $\xi_{2,3}=4$.

Suppose for instance that $\xi_2<4$ and $\xi_3>4$. The resulting
potential is then $V_2$ only and $\xi_2$ slides to zero. In this
limit,
$$
V= -\,{1\over stu\kappa^4},
$$
and the dynamics of $\phi \equiv -\log(stu)$ is described by the
effective Lagrangian
$$
{\cal L}_{\rm eff} = -{e\over2\kappa^2}\left[R
+{1\over6}(\partial_\mu\phi)^2 - {2\over\kappa^2}e^{\phi}\right].
$$
Other scalar components $\log(t/u)$ and $\log(s/u)$ have only
derivative couplings, since the potential only depends on $\phi$.
They can be taken to be constant and arbitrary. The dynamics only
restricts the temperature radius $\kappa^{-2}R^2=e^{-\phi}$, $R_6$
and the string coupling are not constrained, besides inequalities
(\ref{IIphases1}).

In conformally flat gravity background, the equation of motion of
the scalar $\phi$ is
$$
\hat{\Box}\phi =-{6\over\kappa^2}e^\phi.
$$
The solution of the above and the Einstein equations defines a
non-trivial gravitational $\phi$-background. This solution will
correspond to the high-temperature type II vacuum. We will not study
this solution further here. Instead, we will examine in detail in
Sections \ref{sechighphase} and \ref{sechighphase2} the
high-temperature heterotic phase.

\subsection{Five- and six-dimensional limits}

Since we have constructed the effective theory of five-dimensional
strings at finite temperature, an appropriate large radius limit
should lead to a six-dimensional theory at finite temperature. There
should also be a small radius limit leading to a six-dimensional
theory at finite temperature, since torus compactification implies a
radius inversion symmetry. These decompactification limits should,
however, be distinguished from those on $R_6$ which are taken with
fixed four-dimensional coupling $g_H$.

The large radius $R_6$, type IIA limit keeps the temperature radius
$R$ and the five-dimensional coupling
$$
g_5^2 = R_6 g_H^2,
$$
fixed. It corresponds to
\beq
t \rightarrow \infty,\qquad\qquad tu\quad{\rm and}\quad t/s \quad{\rm
fixed}.\label{iialim}
\eeq
On the other hand, the type IIB, small $R_6$ limit
keeps $R$ and the coupling
$$
g_5^2 = {\alpha^\prime_H\over R_6}g_H^2
$$
fixed. It corresponds to
\beq
u \rightarrow \infty,\qquad\qquad tu\quad{\rm and}\quad u/s \quad{\rm
fixed}.\label{iiblim}
\eeq
In both limits, the inequality that separates heterotic
and type II instabilities is
\beq
\label{g5lim}
g_5^2 < \sqrt{\alpha^\prime_H\over2} {\sqrt2+1\over2}.
\eeq
This relation is similar to Eq.(\ref{glim1}) and follows directly
from inequalities (\ref{R6ineq}). The analysis of the
five-dimensional finite-temperature mass formula has been done in
Section \ref{secdual}, in terms of the six-dimensional string
coupling $\lambda_H$. Inequality (\ref{g5lim}) is indeed equivalent
to the bound (\ref{crit}), by simply defining the dimensionless
$\lambda_H$, as in Eq.(\ref{g5is}):
\beq
\lambda_H^2 = g_5^2{R\over\alpha^\prime_H} < {\sqrt2+1\over2}
{R\over\sqrt{2\alpha^\prime_H}} < \left({\sqrt2+1\over2}\right)^2,
\label{hetpha}
\eeq
in both type IIA and IIB theories.

As mentioned in Section 3, the above type IIB theory is defined by
T-duality from the heterotic side. It differs from the type IIB
theory obtained by a T-duality from type IIA [see Eq.(\ref{Tdual})].
The five-dimensional limit of the latter corresponds to a limit where
the heterotic string coupling $\lambda_H$ goes to infinity. This
follows from the duality relations (\ref{sdual}) and (\ref{Tdual}) in
the type IIB decompactification limit
$$
R_6^B\to\infty\qquad {\rm with} \qquad
\lambda_{IIB}\qquad {\rm and} \qquad \alpha^\prime_{II} \qquad {\rm
fixed.}
$$
This takes us outside the bound (\ref{hetpha}), where the non-trivial
high-temperature heterotic phase is defined. A separate analysis is
then needed, which is beyond the scope of this work.

An alternative type IIB theory can, however, be defined directly in
five dimensions by a T-duality from type IIA, reversing the
temperature radius:
$$
R\to {\alpha^\prime_{II}\over R},\qquad
\lambda_{II}\to \lambda_{II}{\sqrt{\alpha^\prime_{II}}\over R}\, .
$$
The high-temperature heterotic phase fixes $R=\sqrt{\alpha^\prime_H}$
or $R_B=\sqrt{\alpha^\prime_{II}}\lambda_{IIA}$ in type IIA units. In
the above type IIB units, this corresponds to $\lambda_{B}=1$, while
the temperature radius remains undetermined. However, in order to
remain in the high temperature heterotic phase, the bound
(\ref{hetpha}) implies
\beq
R_B< {\sqrt{2}+1\over 2}\sqrt{\alpha^\prime_{II}}\, .
\label{boundB}
\eeq

In the high-temperature heterotic phase, valid in the region
(\ref{hetpha}), the temperature is fixed in heterotic units:
\beq
R=\sqrt{\alpha^\prime_H}=
{\sqrt{\alpha^\prime_{II}}\lambda_{IIA}}=
{\alpha^\prime_{II}\over R_B}, \qquad\qquad
\lambda_{H}={1\over\lambda_{IIA}}={\sqrt{\alpha^\prime_{II}}\over R}
={R_B\over\sqrt{\alpha^\prime_{II}}}, \qquad \lambda_B=1\, .
\label{htrel}
\eeq
Thus, in this phase, the {\it only way} to change the temperature is
by varying the heterotic string tension. In particular, the infinite
temperature limit $R\to 0$ is defined by $\alpha^\prime_H\to 0$,
which corresponds to the zero slope field-theory limit of the
corresponding string vacuum. As we will see in the next sections, the
latter is described by a non-critical superstring in six dimensions,
whose zero-slope limit contains a finite number of $N_4=2$ massless
hypermultiplets. This result supports the conjecture that the
high-temperature phase is described by a topological theory
\cite{AW}. From the type IIA side, one may in principle take the
infinite temperature limit by keeping the string tension fixed and
sending its coupling to zero. However, this correspond to
$\lambda_H\to\infty$, which lies outside the domain of validity of
the new phase.

Another interesting limit is the infinite type IIB temperature
$R_B\to 0$, with its string tension fixed. From Eq.(\ref{htrel}),
this corresponds to a zero temperature heterotic theory
($R\to\infty$) with vanishing tension and zero coupling but keeping
the product $\lambda_H^2\alpha^\prime_H$ fixed. Notice the similarity
of this limit to the large-N limit in Yang--Mills theory, with the
Regge slope playing the role of the effective number of degrees of
freedom. This is a non-trivial limit since all genera in principle
contribute. We will return to the above limits in Section 6.

\section{Analysis of the high-temperature heterotic \hfil\break
phase}\label{sechighphase}
\setcounter{equation}{0}

The thermal phase relevant to weakly coupled, high-temperature
heterotic strings at intermediate values of the radius $R_6$ [see
inequalities (\ref{glim1}) and (\ref{R6ineq})] has an interesting
interpretation; we study this here, using the information contained
in its effective theory, which is characterized by
Eqs.(\ref{highhetmin}):
\beq
\label{hethighvac}
tu = 1, \qquad H_1 = {1\over2}, \qquad H_2 = H_3 = 0.
\eeq
These values solve the equations of motion of all scalar
fields with the exception of $s=\Re S$.
The resulting bosonic effective Lagrangian describing the dynamics of
$s$ and $g_{\mu\nu}$ is
\beq
\label{hethigh1}
{\cal L}_{\rm bos} = -{1\over2\kappa^2}eR - {e\over4\kappa^2}
(\partial_\mu\ln s)^2 + {e\over2\kappa^4s}.
\eeq
For all (fixed) values of $s$, the cosmological constant is negative
since $V=-(2\kappa^4s)^{-1}$ and the apparent
geometry is anti-de Sitter.
But the effective theory (\ref{hethighvac}) does not stabilize $s$.

To study the bosonic Lagrangian, we first rewrite it in the string
frame. Defining the dilaton as
\beq
\label{dilis}
e^{-2\phi} = s,
\eeq
and rescaling the metric according to
\beq
\label{grescal}
g_{\mu\nu} \quad\longrightarrow\quad
{2\kappa^2\over\alpha^\prime_H}e^{-2\phi} g_{\mu\nu},
\eeq
one obtains\footnote{Since the rescaling $g_{\mu\nu}\rightarrow
e^{-2\sigma}g_{\mu\nu}$ leads to
$e[R+6(\partial_\mu\sigma)^2]\rightarrow
e^{-2\sigma}eR$.}
\beq
\label{stframe}
{\cal L}_{\rm string~frame} = {e^{-2\phi}\over\alpha^\prime_H}
\left[-eR+4e(\partial_\mu\phi)(\partial^\mu\phi) +
{2e\over \alpha^\prime_H} \right].
\eeq
The equation of motion for the dilaton then is
\beq
\label{dileom}
R +4(\partial_\mu\phi)(\partial^\mu\phi) -4\Box\phi =
{2\over\alpha^\prime_H}.
\eeq
Comparing with the two-dimensional sigma-model dilaton
$\beta$-function \cite{beta} with central charge deficit
$\delta c = D-26$, which leads to
\beq
\label{dileom2}
R +4(\partial_\mu\phi)(\partial^\mu\phi) -4\Box\phi =
- {\delta c\over3\alpha^\prime_H},
\eeq
we find a central charge deficit $\delta c = -6$, or, for a
superstring\footnote{The same analysis in Ref.\cite{AK} is in error
by a factor 2.},
\beq
\label{deltac}
\delta\hat c = {2\over3}\delta c = -4.
\eeq

In the string frame, a background for theory (\ref{stframe}) has flat
(sigma-model) metric\footnote{The notation $\,\,\tilde{}\,\,$ is used
for a background field.} ${\tilde g}_{\mu\nu}=\eta_{\mu\nu}$ and
linear dilaton dependence \cite{ABEN} on a spatial coordinate, say
$x^1$:
\beq
\label{lindil}
\tilde\phi = Q x^1, \qquad\qquad
Q^2 = {\delta\hat c\over 8\alpha^\prime_H}={1\over2\alpha^\prime_H}.
\eeq
In the flat background, the Lagrangian density for the dilaton
expanded up to quadratic order in $\varphi=\phi-Qx^1$ is
$$
{\cal L}_{dil.} = {16Q\over\alpha^\prime_H}e^{-2Qx^1}
\varphi(\partial_1\varphi)
+{4\over\alpha^\prime_H}e^{-2Qx^1}
(\partial_\mu\varphi)(\partial^\mu\varphi)
-{8\over\alpha^{\prime2}_H}e^{-2Qx^1}\varphi^2,
$$
omitting a $\varphi$--independent contribution.
Defining then the rescaled field
$$
\hat\varphi = \varphi\, e^{-Qx^1}\,,
$$
one obtains the equivalent Lagrangian
\beq
\label{dilmass}
{\cal L}_{\rm dil} = {4\over\alpha^\prime_H}\left[
(\partial_\mu\hat\varphi)(\partial^\mu\hat\varphi)
+{1\over2\alpha^\prime_H}\hat\varphi^2 \right],
\eeq
which indicates that a scalar field with mass
\beq
\label{dilmass2}
m_{\rm dil}^2 = {1\over2\alpha^\prime_H}= Q^2
\eeq
propagates in the background.

A similar analysis can be applied to the axionic partner $\Im S=a$ of
the supergravity dilaton $s=\Re S$. Its (bosonic) Lagrangian is
simply
$$
{\cal L}_a = -{e\over4s^2\kappa^2}(\partial^\mu a)(\partial_\mu a),
$$
in the Einstein frame and
$$
{\cal L}_{a, {\rm string}} = -{e\over2\alpha^\prime_H}
e^{-2\phi}(\partial^\mu a)
(\partial_\mu a)
$$
in the string frame, according to rescaling (\ref{grescal}). In the
linear dilaton background $\phi=\tilde\phi=Qx^1$, the rescaled axion
$\hat a = e^{-Qx^1}a$ has quadratic Lagrangian
$$
-{1\over2\alpha^\prime_H}\left[ (\partial^\mu\hat a)
(\partial_\mu\hat a) +Q^2\hat a^2 \right],
$$
and its mass squared is again $Q^2$. The same mass shift by the
quantity $Q^2=m_{3/2}^2$ will appear in all scalar masses computed in
the linear dilaton background \cite{ABEN}.

Before turning to the complete analysis of the mass spectrum in the
high-tempe\-ra\-tu\-re heterotic phase, we now establish the residual
supersymmetries expected in the background chracterized by the linear
dilaton dependence on $x^1$.

\subsection{Broken supersymmetry}

The linear dilaton background breaks both four-dimensional Lorentz
symmetry and four-dimensional Poincar\'e supersymmetry. Since
supersymmetry breaks sponta\-neous\-ly, one expects to find goldstino
states in the fermionic mass spectrum and massive spin 3/2 states.
And, because of the non-trivial background, the theory in the
high-temperature heterotic phase is effectively a three-dimensional
supergravity.

Local supersymmetry in three-dimensional space-time is notoriously
difficult to establish in the presence of massive states. This is
because \cite{W, BBS} masses induce asymptotically a conical geometry
\cite{3dgrav}, making Noether supercharges hard to define.
Supersymmetry of the vacuum does not necessarily imply supersymmetry
of the massive spectrum. This phenomenon exists in (locally) flat
background and the presence of the linear dilaton does not simplify
the matter. In the following paragraphs, we will first consider the
existence of goldstino fermions in the high-temperature vacuum, then
compute the mass spectrum, which will turn out to be supersymmetric
for moduli $T$ and $U$ (which are not massless) and perturbative
heterotic windings $Z_1^\pm$. This supersymmetry in the spectrum will
however be broken in the non-perturbative sector $Z_2^\pm$ and
$Z_3^\pm$. Finally, by taking the five-dimensional limit discussed in
the previous section, we will observe that this non-perturbative
breaking of supersymmetry persists, indicating clearly, in five
dimensions, broken supersymmetry.

To discuss the pattern of goldstino states, observe first that
the supergravity extension of the bosonic Lagrangian
(\ref{hethigh1}) includes a non-zero gravitino mass
term for all values of $s$ since
\beq
\label{m3/2}
m_{3/2}^2 = \kappa^{-2}\, e^{\cal G} = {1\over4\kappa^2 s} =
{1\over2\alpha^\prime_H} = Q^2.
\eeq
Notice also for future use that the potential at the vacuum
verifies
\beq
\label{potvac}
V = -{2\over\kappa^4}e^{\cal G} = -{1\over2\kappa^4 s}
= -{2\over \kappa^2}\,m_{3/2}^2.
\eeq
Then, consider the transformation of fermions in the chiral multiplet
$(z^i,\chi^i)$ \footnote{The notation is as in Ref.\cite{CFGVP},
with sign-reversed ${\cal G}$ and $\sigma^{\mu\nu} =
{1\over4}[\gamma^\mu,\gamma^\nu]$. Indices $i,j,\ldots$, enumerate
all chiral multiplets $(z^i,\chi^i)$.}:
\beq
\label{susytransf}
\delta\chi_{Li} = {1\over2}\kappa (\slash\partial z_i)\epsilon_R
-{1\over2}e^{{\cal G}/2}\,({\cal G}^{-1})_i^j{\cal G}_j \,\epsilon_L
+\ldots,
\eeq
omitting fermion contributions. In the high-temperature heterotic
phase,
\beq
\label{Gmin}
{\cal G}_S= {\partial\over\partial S}{\cal G} = -{1\over2s},
\qquad\qquad
{\cal G}_a = {\partial\over\partial z^a}{\cal G} =0,
\eeq
and the K\"ahler metric is diagonal with
${\cal G}^S_S = (2s)^{-2}$. Since also
$$
\slash\partial s = -2Q s\gamma^1, \qquad e^{{\cal G}/2} = \kappa Q,
$$
only the fermionic partner $\chi_s$ of the dilaton $s$
participates in supersymmetry breaking, with the transformation
\beq
\label{susytranf2}
\delta \chi_s = {\sqrt s\over2}(1-\gamma^1)\epsilon.
\eeq
Supersymmetries generated by $(1-\gamma^1)\epsilon$ are then broken
in the linear dilaton background in the $x_1$ direction while those
with parameters $(1+\gamma^1)\epsilon$ remain unbroken. Starting then
from sixteen supercharges ($N_4=4$ supersymmetry) at zero
temperature, the high-temperature heterotic vacuum has eight unbroken
supercharges. Since the effective space-time symmetry is
three-dimensional, the high-temperature phase has $N_3=4$
supersymmetry: the linear dilaton background acts identically with
respect to the $N_4=4$ spinorial charges. It simply breaks one half
of the charges in each spinor. Thus, the high-temperature phase is
expected to be stable because of supersymmetry of its effective field
theory and because of its superconformal content\footnote{See next
section.}.

The pattern of supersymmetry breaking can be confirmed by computing
the gra\-vi\-ti\-no--$\chi_s$ quadratic couplings, which generate the
super-Higgs phenomenon. In the linear dilaton background, the
non-kinetic quadratic fermionic terms are:
\beq
\label{lindilferm}
e^{-1}{\cal L}_{3/2-1/2} =
-Q\, \ov\psi_\mu\sigma^{\mu\nu}\gamma_5\psi_\nu
-{Q\over2s}\,\ov\psi_\mu(1+\gamma^1)\gamma^\mu\gamma_5\chi_s
-{3\over2}{Q\over(2s)^2}\,\ov\chi_s\gamma^1\gamma_5\chi_s.
\eeq
We then separate the three-dimensional gravitino $\psi_m$, $m=0,2,3$
from the spinor $\psi_1$, with the redefinition
$$
\psi_m +{1\over2}\gamma_m\gamma^1\psi_1 \quad\longrightarrow\quad
\psi_m,
$$
and the gravitino contributions become
$$
-Q\,\ov\psi_m\sigma^{mn}\gamma_5\psi_n
-{Q\over2s}\, \ov\psi_m\gamma^m(1-\gamma^1)\gamma_5\chi_s.
$$
These terms identify the goldstino fermion as
\beq
\label{Gold}
\psi_G \sim {1\over2s}(1-\gamma^1)\gamma_5\chi_s,
\eeq
in agreement with the result (\ref{susytranf2}), which indicates that
supersymmetries with parameter $(1-\gamma^1)\epsilon$ are broken.

At the level of the background solution, one would conclude that one
half of the supersymmetries remain unbroken ($N_3=4$). We now
confront this statement with the mass spectrum in the scalar and spin
1/2 sectors of the effective supergravity.

\subsection{Mass spectrum}\label{secmass}

We now analyse the complete mass spectrum of the effective
supergravity theory in the linear dilaton background relevant to the
high-temperature heterotic phase. This spectrum naturally splits in
two sectors. First, as already discussed in the previous paragraph,
the heterotic dilaton multiplet, with scalar $S$ and spin 1/2 partner
$\chi_s$, is actively involved in the fate of supersymmetry in the
high-temperature phase and in the background. Secondly, all other
chiral multiplets play a passive role in these respects. To simplify
the notation, we will in this paragraph collectively denote the
scalar fields $T$, $U$, $Z_A^+$ and $Z_A^-$ by $y_a$, and their spin
1/2 partners by $\chi_a$.

This splitting of the chiral multiplets arises because in the
`vacuum' defined by Eqs.(\ref{hethighvac}), the K\"ahler function
${\cal G}$ does not induce any mixing of $(S,\chi_s)$ with
$(y_a,\chi_a)$ [see Eqs.(\ref{Gmin})] and
supersymmetry breaking is entirely decided in the $S$ sector.
The K\"ahler metric is diagonal,
${\cal G}^S_a = {\partial^2{\cal G}\over\partial S^*\partial z^a}=
0$,
for all values of the fields. In addition,
$$
{\cal G}_{Sa} = {\partial^2\over\partial S\partial y^a}{\cal G} = 0.
$$
Notice, however, that ${\cal G}_{Sab}= {\partial^3{\cal
G}\over\partial S \partial y^a\partial y^b}$ does not vanish: it will
generate a contribution to the mass spectrum in the $y^a$ sector.

The splitting of $S$ and $y^a$ does not exist in the
low-temperature phase $H_1=H_2=H_3=0$, in which
\beq
\label{lowphase1}
{\cal G}_S=-(2s)^{-1},\qquad
{\cal G}_T=-(2t)^{-1},\qquad
{\cal G}_U=-(2u)^{-1},
\eeq
with
$$
\psi_G = {1\over2s}\chi_s + {1\over2t}\chi_t + {1\over2u}\chi_u
$$
as goldstino state\footnote{Expressed using non-normalized fermions.
Canonical normalization of the spinors would lead to
$\psi_G=\chi_s+\chi_t+\chi_u$.}. The low-temperature phase is
symmetric in the moduli $s$, $t$ and $u$: it is common to the three
dual strings, in their perturbative and non-perturbative domains. In
contrast, the high-temperature heterotic phase only exists in the
perturbative domain of the heterotic string, where $s$ is the
dilaton, and, by duality, in non-perturbative type II regimes.

We have already considered the gravitino states and fields of the
chiral multiplet $(S=s+ia,\chi_s)$. We now turn to the fermions
$\chi_a$ and the scalars $y_a$. It is useful to recall that the
K\"ahler metric in the high-temperature phase is diagonal and
particularly simple:
\beq
\label{metric}
\begin{array}{c}
{\cal G}_S^S = (2s)^{-2}, \qquad
{\cal G}_T^T = (2t)^{-2}, \qquad
{\cal G}_U^U = (2u)^{-2}, \qquad (tu=1), \crbig
\displaystyle{{\partial^2{\cal G}\over\partial Z_A^\pm\partial
Z_B^{\pm *}}
= {2\over (1-z_1^2)^2}\delta_{AB}
= {1\over2z_1^2}\delta_{AB},
\qquad (A,B=1,2,3).}
\end{array}
\eeq
The last equality follows from $H_1=z_1/(1-z_1^2)=1/2$.
These results will be used to canonically normalize massive fields.

\vspace{3mm}
\noindent{$\bullet$\,\, \it Fermions $\chi_a$:}

\noindent
The mass matrix of the spin 1/2 partners of $T$, $U$, $Z_A^+$ and
$Z_A^-$, with inverse K\"ahler metric factors to canonically
normalize fields, is simply
$$
({\cal M}_{1/2})_{ab} = \kappa^{-1} \,e^{{\cal G}/2}\, ({\cal
G}^{-1/2})^c_a
\left( {\cal G}_{cd} + {1\over3}{\cal G}_c{\cal G}_d-
{\cal G}_e{{\cal G}^{-1}}_f^e{\cal G}^f_{cd}\right)({\cal
G}^{-1/2})^d_b.
$$
Since the K\"ahler metric is diagonal, ${\cal G}_b=0$ and
${\cal G}^S_{cb}=0$, the mass matrix simplifies to
\beq
\label{M1/21}
({\cal M}_{1/2})_{ab} = \kappa^{-1}\,e^{{\cal G}/2}\,
({\cal G}^{-1/2})^c_a {\cal G}_{cd}({\cal G}^{-1/2})^d_b =
m_{3/2}\,({\cal G}^{-1/2})^c_a {\cal G}_{cd}({\cal G}^{-1/2})^d_b.
\eeq
Mixings can only arise from non-zero values of ${\cal G}_{ab}$ due to
superpotential contributions. Since $W$ includes a term proportional
to $TUZ_1^+Z_1^-$, these four fields, which are non-zero at the
vacuum, will get mixed. Masses will be completely determined (in
$m_{3/2}$ units) since all parameters are fixed in this sector. On
the other hand, the two fermion masses in the $Z_2^\pm$ sector are
$m_{3/2}[su\pm 1]$ and $m_{3/2}[st\pm1]$ in the $Z_3^\pm$ sector.

\vspace{3mm}
\noindent{$\bullet$\,\, \it Scalars $y_a$:}

\noindent
The high-temperature `vacuum' is a minimum of the potential for the
scalars $y^a$, $V_a={\partial V\over\partial y^a}=0$, and also
${\cal G}_a = 0$, according to Eq.(\ref{Gmin}).
Again, since
$$
{\partial^2 V\over\partial S\partial y^a} =
{\partial^2 V\over\partial S\partial y^*_a} = 0,
$$
at the minimum, the scalar mass matrix splits into mass terms for $S$
and a mass matrix for the scalars $y^a$, which is given by
$$
{\cal M}_0^2 = \kappa^{-2}
\left([{\cal G}^{-1/2}]^a_e \,\, [{\cal G}^{-1/2}]^f_c\right)
\left(\begin{array}{cc} V^e_g & V^{eh} \crbig
V_{fg} &  V^h_f \end{array}\right)
\left( \begin{array}{c}
[{\cal G}^{-1/2}]^g_b  \crbig
[{\cal G}^{-1/2}]^d_h \end{array} \right).
$$
The metric factors $[{\cal G}^{-1/2}]$ normalize the fields
canonically. Each term can be computed at the high-temperature vacuum
and one obtains
$$
\begin{array}{rcl}
{\cal M}_0^2 &=&\displaystyle{
m_{3/2}^2
\left([{\cal G}^{-1/2}]^a_e \,\, [{\cal G}^{-1/2}]^f_c\right)
\left(\begin{array}{cc}
{\cal G}^{en} {{\cal G}^{-1}}^r_n {\cal G}_{rg} &
-2sW^{-1}W^{ehS} \crbig
-2sW^{-1}W_{fgS} &
{\cal G}_{fm} {{\cal G}^{-1}}^m_p {\cal G}^{ph}
\end{array}\right)
\left( \begin{array}{c}
[{\cal G}^{-1/2}]^g_b  \crbig
[{\cal G}^{-1/2}]^d_h \end{array} \right)
}\crbig
&& -m_{3/2}^2\left(\begin{array}{cc} \delta^a_b & 0 \crbig 0 &
\delta^d_c
\end{array}\right).
\end{array}
$$
As already observed, the linear dilaton background further shifts
this mass matrix by a quantity that precisely cancels the last
contribution. The physical scalar mass matrix then becomes
\beq
\label{M01}
\begin{array}{rcl}
{\cal M}_0^2 &=&\displaystyle{
m_{3/2}^2
\left([{\cal G}^{-1/2}]^a_e \,\, [{\cal G}^{-1/2}]^f_c\right)
\left(\begin{array}{cc}
{\cal G}^{en} {{\cal G}^{-1}}^r_n {\cal G}_{rg} &
-2sW^{-1}W^{ehS} \crbig
-2sW^{-1}W_{fgS} &
{\cal G}_{fm} {{\cal G}^{-1}}^m_p {\cal G}^{ph}
\end{array}\right)
\left( \begin{array}{c}
[{\cal G}^{-1/2}]^g_b  \crbig
[{\cal G}^{-1/2}]^d_h \end{array} \right).}
\end{array}
\eeq
Comparing with the fermion mass matrix (\ref{M1/21}), one observes
that the spectrum would be supersymmetric\footnote{In the sense of
equal boson and fermion masses.} without the off-diagonal term
proportional to $2sW^{-1}W^{Sij}$. Since supersymmetry breaks in the
$S$ direction, these off-diagonal contributions generate
O'Raifeartaigh-type bosonic mass shifts for states that couple in the
superpotential to $S$: these are the heterotic dyonic states
$Z_2^\pm$ and $Z_3^\pm$, which generate type II thermal
instabilities. As observed in Ref.\cite{AK}, heterotic perturbative
states have a supersymmmetric spectrum.

These supersymmetry-breaking contributions to the scalar mass
spectrum imply the existence of non-perturbative modes lighter than
their fermionic partners. Explicitly, since
$$
-2sW^{-1}{\partial^3 W\over\partial S\partial Z_2^+\partial Z_2^-} =
-{su\over z_1^2}, \qquad
-2sW^{-1}{\partial^3 W\over\partial S\partial Z_3^+\partial Z_3^-} =
-{st\over z_1^2}, \qquad (t=1/u),
$$
the scalar mass matrix in the $Z_2$ sector reads
$$
\begin{array}{r}
Z_2^+: \crbig Z_2^-: \crbig Z_2^{+*}: \crbig Z_2^{-*} :
\end{array} \qquad
m^2_{3/2}\left(\begin{array}{cccc}
(su)^2+1 & -2su & 0 & -2su \crbig
-2su & (su)^2+1 & -2su & 0 \crbig
0 & -2su & (su)^2+1 & -2su \crbig
-2su & 0 & -2su & (su)^2+1
\end{array}\right).
$$
The eigenvalues are
$$
m^2_{3/2}[(su-1)^2 \pm 2su], \qquad\qquad
m^2_{3/2}[(su+1)^2 \pm 2su],
$$
to be compared with the fermion masses $|su-1|m_{3/2}$ and
$|su+1|m_{3/2}$. The mass pattern in the $Z_3^\pm$ sector is obtained
by substituting $u$ for $t$ in the $Z_2^\pm$ sector.

To summarize, the spectrum is supersymmetric in the perturbative
heterotic and moduli sector ($T,U,Z_1^\pm$), and with O'Raifeartaigh
pattern in the non-perturbative sectors:
$$
\begin{array}{rrcl}
Z_2^\pm:&  m_{bosons}^2 &=& m_{fermions}^2 \pm 2su\, m_{3/2}^2,
\crbig
Z_3^\pm:&  m_{bosons}^2 &=& m_{fermions}^2 \pm 2st\, m_{3/2}^2.
\end{array}
$$

This phenomenon persists in the five-dimensional type IIA [and type
IIB] limit (\ref{iialim}) [and (\ref{iiblim})], in which the
$Z_3^\pm$ [$Z_2^\pm$] states become superheavy and decouple while the
$Z_2^\pm$ [$Z_3^\pm$] scalar masses are shifted by a non-perturbative
amount, since in this limit $su=1/\lambda_H^2$ [$st=1/\lambda_H^2$].
Thus, supersymmetry appears to be broken by non-perturbative effects.
Note again, however, that this statement may not hold in the case of
the four-dimensional background solution with a dilaton motion in one
direction. In this case, there is only an effective three-dimensional
Poincar\'e invariance, which {\it does not} imply in general mass
degeneracy within a massive multiplet, even if local supersymmetry is
unbroken \cite{W, BBS}.

In the special infinite heterotic temperature limit discussed at the
end of Section \ref{seceff}, where $\alpha^\prime_H\to 0$, all
massive states decouple and consequently one recovers $N_4=2$
unbroken (rigid) supersymmetry in the effective (topological) field
theory of the remaining massless hypermultiplets.

\section{The high-T Heterotic Phase Transition}\label{sechighphase2}
\setcounter{equation}{0}

As we discussed in Sections \ref{seceff} and \ref{sechighphase}, the
high-temperature phase of $N_4=4$ strings is described by a
non-critical string with central charge deficit $\delta {\hat c}=-4$,
provided the (six-dimensional) heterotic string is in the weakly
coupled regime with $\lambda_H\le ({\sqrt 2}+1)/2$. One possible
description is in terms of the (5+1) super-Liouville theory
compactified (at least) on the temperature circle with radius fixed
at the fermionic point $R=\sqrt{\alpha^\prime_H}$. The perturbative
stability of this ground state is guaranteed when there is at least
$\rm {\cal N}_{sc}=2$ superconformal symmetry on the world-sheet,
implying at least $N_4=1$ supersymmetry in space-time. However, our
analysis of the previous section shows that the boson--fermion
degeneracy is lost at the non-perturbative level, although the ground
state remains supersymmetric.

An explicit example with $\rm {\cal N}_{sc}=4$ superconformal was
given in Ref.\cite{K, AFK}. It is obtained when together with the
temperature circle there is an additional compactified coordinate on
$S^1$ with radius $R_6=\sqrt{\alpha^\prime_H}$. These two circles are
equivalent to a compactification on $[SU(2)\times SU(2)]_k$ at the
limiting value of level $k=0$. Indeed, at $k=0$, only the 6
world-sheet fermionic $SU(2)\times SU(2)$ coordinates survive,
describing a ${\hat c}=2$ system instead of ${\hat c}=6$ of
$k\to\infty$, consistently with the decoupling of four
supercoordinates, $\delta {\hat c}=-4$. The central charge deficit is
compensated by the linear motion of the dilaton associated to the
Liouville field, $\phi=Q^\mu x_\mu$ with $Q^2=1/(2\alpha^\prime_H)$
so that $\delta {\hat c}_L=8\alpha^\prime_H~Q^2=4$. Using the
techniques developed in Refs.\cite{ABK, AFK}, one can derive the
one-loop (perturbative) partition function in terms of the left- and
right-moving degrees of freedom on the world-sheet. Namely:

\begin{enumerate}
\item
The four left- and right-moving non-compact coordinates $x_\mu$
(which include the Liouville coordinate) together with the
reparametrization ghosts $b,c;{\ov b}, {\ov c}$. Their contribution
to the partition function is:
\beq
Z\{x_{\mu},b,c;\,{\ov b}, {\ov c}\}~=
{}~{{\rm Im}\tau^{-1}\over
\eta^2~{\ov\eta}^{2}} \,;
\eeq
\item
The two left- and right-moving coordinates $\phi_1,~\phi_2$
compactified on $S^1_R\times S^1_{R_6}$ at the fermionic point,
$R=R_6=\sqrt{\alpha^\prime_H}$. By fermionization the currents
$\partial\phi_1$, $\partial\phi_2$ and ${\ov\partial}\phi_1$,
${\ov\partial}\phi_2$ are equivalent to four left- and four
right-moving world-sheet fermions $\chi_I$, and $\ov{\chi}_I,~
(I=1,2,3,4) $ giving rise to an $SO(4)_{\rm left}\times SO(4)_{\rm
right }$ current algebra. Their contribution to the partition
function is given in terms of the $SO(4)_{\rm left}\times SO(4)_{\rm
right }$ characters:
\beq
Z\{\chi_I,~ \ov{\chi}_I\}={1\over \eta^2~{\ov\eta}^2}
\theta^2\left[^{~\alpha + h~}_{~\beta + g~} \right]~
{\ov\theta}^2\left[^{~{\ov \alpha} + h~}_{~{\ov \beta} + g~}
\right]\, ,
\eeq
where the spin structures $\alpha$, $\beta$, ${\ov \alpha}$, ${\ov
\beta}$, $h$ and $g$ take values 0 or 1.
\item
The remaining left-moving degrees of freedom $\Psi_{\mu}$,
$\mu=1,\ldots,6$, are the superpartners of the coordinates
$x_\mu,\phi_1, \phi_2$, and the super-reparametrization ghosts
$\beta,~\gamma$. Their partition function reads:
\beq
Z\{\Psi_{\mu},\beta,\gamma\}~=~{1\over
\eta^4}~\theta^2\left[^{~\alpha~}_{~\beta ~} \right].
\eeq
\item
The right-moving degrees of freedom also include a conformal system,
which can be described by 28 fermions ${\ov \Psi}_A$, with central
charge:
$$
c_R[{\ov \Psi_A}]~=~26-~4~({\rm from}~~ x_{\mu})-
{}~2~({\rm from}~~ \phi_1,\phi_2)-
{}~6~ \left({\rm from}~~
{3\over2}\delta{\hat c}\right)~=~14\,.
$$
Their contribution to the partition function in terms
of $SO(28)$ characters is:
\beq
Z\{{\ov \Psi_A}\}~=~{1\over {\ov \eta}^{14}}~{\ov
\theta}^{\,14}\left[^{~{\ov \alpha}~}_{~{\ov \beta} ~} \right].
\eeq
\end{enumerate}
Assembling the above conformal blocks, one obtains the partition
function of the $(5+1)$-dimensional Liouville model, with the desired
${\rm \cal N}_{sc}=4$ superconformal symmetry:
\beq
\label{pf}
\begin{array}{l}
\displaystyle{Z^{\rm Liouv}[SU(2)\times SU(2)]_{k=0}~=~ {{\rm
Im}\tau^{-1}\over
\eta^6~{\ov\eta}^{18}} ~{1\over 8}
\sum_{\alpha,\beta,{\ov \alpha},{\ov\beta},h,g}
(-)^{\alpha+\beta+\alpha\beta}} \crbig
\displaystyle{\times~\theta^{~2}\left[^{~\alpha~}_{~\beta~} \right]
{}~\theta\left[^{~\alpha + h~}_{~\beta + g~} \right]~
\theta\left[^{~\alpha - h~}_{~\beta - g~} \right]
{}~~{\ov\theta}\left[^{~{\ov\alpha} +h~}_{~{\ov\beta}+g~} \right]
{}~{\ov\theta}\left[^{~{\ov\alpha} -h~}_{~{\ov\beta} -g~} \right]
{}~{\ov\theta}^{~14}\left[^{~{\ov\alpha}~}_{~{\ov\beta}~}
\right]}.
\end{array}
\eeq
This partition function encodes a number of properties, which
deserve some comments:
\begin{itemize}
\item
The initial $N_4=4$ supersymmetry is reduced to $N_4=2$ (or $N_3=4$)
because of the $Z_2$ projection generated by $(h,g)$. This agrees
with our effective field theory analysis of the high-temperature
phase given in Section \ref{sechighphase}. The (perturbative) bosonic
and fermionic mass fluctuations are degenerate due to the remaining
$N_4=2$ supersymmetry.
\item
The $h=0$ sector does not have any massless fluctuations due to the
linear dilaton background or to the temperature coupling. The linear
dilaton background shift the bosonic masses (squared) by $m^2_{3/2}$,
so that all bosons in this sector have masses larger than or equal to
$m_{3/2}$. This is again in agreement with our effective theory
analysis. Similarly, fermion masses are shifted by the same amount
because of the $S^1_R$ temperature modification.
\item
In the $h=1$ ``twisted", sector there are massless excitations as
expected from the (5+1) super-Liouville theory \cite{BG, KS, AFK}.
\item
The $5+1$ Liouville background can be regarded as a Euclidean
five-brane solution wraped on $S^1\times S^1$ preserving one-half of
the space-time supersymmetries ($N_4=2$).
\item
The massless space-time fermions coming from the $h=1$ sector are
six-dimen\-sio\-nal spinors constructed with the $\Psi_{\mu}$ and
$\beta,\gamma$. They are also spinors under the $SO(4)_{\rm right}$
constructed using ${\ov \chi}_I$, and vectors under $SO(28)$
constructed with ${\ov \Psi}_A$.
\item
The massless space-time bosons are in the same right-moving
representation {\it e.g.} $SO(4)_{\rm right}$ spinors and $SO(28)$
vectors. In addition, they are spinors under $SO(4)_{\rm left}$
constructed with ${\chi}_I$. Together with the massless fermions,
they form 28 $N_4=2$ hypermultiplets.
\end{itemize}
These 28 massless hypermultiplets are the only states that survive in
the zero-slope limit and their effective field theory is described by
a $N_4=2$ sigma-model on a hyper-K\"ahler manifold. This topological
theory corresponds to the infinite temperature limit of the $N_4=4$
strings after the heterotic Haggedorn phase transition.

Although the $5+1$ Liouville background is perturbatively stable due
to the ${\rm \cal N}_{sc}=4$ superconformal symmetry, its stability
is not ensured at the non-perturbative level when the heterotic
coupling is large:
\beq
g^2_H(x_{\mu})=e^{2(\phi_0 - Q^{\mu}x_{\mu}) }>
{\sqrt{2}+1 \over 2\sqrt{2}} \sim 0.8536.
\eeq
Indeed, as we explained in Paragraph \ref{secphase}, the
high-temperature heterotic phase only exists if $g^2_H(x_{\mu})$ is
lower than a critical value separating the heterotic and Type II
high-temperature phases. Thus one expects a domain wall in
space-time, at $x_{\mu}^0=0$, separating these two phases:
$g^2_H(Q^{\mu} x^0_{\mu})\sim 0.8536$. This domain wall problem can
be avoided by replacing the $(5+1)$ super-Liouville background with a
more appropriate one with the same superconformal properties, $\rm
{\cal N}_{sc}=4$, obeying however the additional perturbative
constraint $g^2_H(x_{\mu})<<1$ in the entire space-time.

Exact superstring solutions based on gauged WZW two-dimensional
models with $\rm {\cal N}_{sc}=4$ superconformal symmetries have been
studied in the literature \cite{ABS, KPR, K, AFK, KKL}. We now
consider the relevant candidates with $\delta \hat{c}=-4$.

The first one is the $5+1$ super-Liouville with $\delta{\hat c}=4$,
already examined above. It is based on the $2d$-current algebra:
\begin{eqnarray}
&&U(1)_{\delta\hat c=4} \times U(1)^3 \times
U(1)_{R^2={\alpha^\prime_H}}
\times U(1)_{R_6^2={\alpha^\prime_H}}
\nonumber \\
&& \equiv ~U(1)_{\delta\hat c=4} \times U(1)^3 \times SO(4)_{k=1}.
\end{eqnarray}

Another class of candidate background is made up of the non-compact
parafermionic spaces described by gauged WZW models:
\beq
\begin{array}{rl}
&\left[{SL(2,R) \over U(1)_{V,A} }\right]_{k=4} \times
\left[{SL(2,R) \over U(1)_{V,A} }\right]_{k=4} \times
 U(1)_{R^2={\alpha^\prime_H}} \times  U(1)_{R_6^2={\alpha^\prime_H}}
\crbig
\equiv& \left[{SL(2,R) \over U(1)_{V,A} }\right]_{k=4} \times
\left[{SL(2,R) \over U(1)_{V,A} }\right]_{k=4}
\times SO(4)_{k=1}\,,
\end{array}
\eeq
where indices $A$ and $B$ stand for the ``axial" and ``vector"
WZW $U(1)$ gaugings.

Then, many backgrounds can be obtained by marginal deformations of
the above, preserving at least $\rm{\cal N}_{sc}=2$, or also by
acting with S- or T-dualities on them.

As already explained, the appropriate background must verify
the weak-coupling constraint:
\beq
\label{lastequ}
g^2_H(x_{\mu})=e^{2\phi}<<\,\sim \,0.8536 \,,
\eeq
in order to avoid the domain-wall problem, and in order to trust the
perturbative validity of the heterotic string background. This
weak-coupling limitation is realized in the ``axial" parafermionic
space. In this background, $g^2_H(x_{\mu})$ is bounded in the entire
non-compact four-dimensional space, with coordinates
$x_{\mu}=\{z,z^*,w,w^*\}$, provided the initial value of
$g^2_0=g^2_H(x_{\mu}=0)$ is small.
\beq
{1\over g^2_H(x_{\mu})}=e^{-2\phi}={1\over g^2_0}~(1+zz*)~(1+ww*)~\ge
{}~{1\over g^2_0}.
\eeq
The metric of this background is everywhere regular:
\beq
ds^2={4dzdz^*\over 1+zz^*} +{4dwdw^*\over 1+ww^*}\,\,.
\eeq
The Ricci tensor
\beq
R_{z\,z^*}={1\over (1+zz^*)^2}~, {\rm~~~~~~} R_{w\,w^*}={1\over
(1+ww^*)^2}\,\,.
\eeq
The scalar curvature
$$
R={1\over 4(1+zz^*)}+{1\over 4(1+ww^*)}
$$
vanishes for asymptotically large values of $|z|$ and $|w|$
(asymptotically flat space). This space has maximal curvature
when $|z|=|w|=0$. This solution has a behaviour similar to that
of the Liouville solution in the asymptotic regime
$|z|,~|w|~\rightarrow ~\infty$. In this limit, the dilaton $\phi$
becomes linear when expressed in terms of the flat
coordinates $x_i$:
\beq
\phi=-\Re[{\rm log}z]-\Re[{\rm log}w] = -Q^1|x_1|-Q^2|x_2|,
\eeq
where
$$
x_1=-{\Re}[{\rm log}z], \quad
x_2=-{\rm Re}[{\rm log}w], \quad
x_3={\rm Im}[{\rm log}z], \quad
x_4={\rm Im}[{\rm log}w],
$$
and the line element is $ds^2=4(dx_i)^2$. The important point here is
that, for large values of $|x_1|$ and $|x_2|$, $\phi\ll0$, in
contrast to the Liouville background in which $\phi=Q^1x_1+Q^2x_2$,
the dilaton becomes positive and arbitrarily large in one half of the
space, violating the weak-coupling constraint (\ref{lastequ}).

We then conclude that the high-temperature phase is described by the
above parafermionic space, which is stable because of $N_4=2$
supersymmetry. Since it is perturbative everywhere, the perturbative
massive bosonic and fermionic fluctuations are always degenerate,
while the non-perturbative ones are superheavy and decouple in the
limit of vanishing coupling.

\section{Conclusions}\label{secconclusions}
\setcounter{equation}{0}

In this work we studied string theory at finite temperature $T$ and
the issue of Hagedorn transition, using the recent non-perturbative
understanding of the theory based on string dualities. For
simplicity, we restricted ourselves to the simplest case of $N_4=4$
compactifications in $D=6$ dimensions, obtained by compactifying the
heterotic string on $T^4$ or the type II string on $K_3$. As usual,
the thermodynamics can be described by introducing an additional
compactification of the (Euclidean) time on a circle of radius
$R=1/(2\pi T)$. In this context, finite temperature boundary
conditions correspond to a particular gauging of the $N_4=4$
supersymmetry, while Hagedorn instabilities of different perturbative
string descriptions appear as thermal dyonic 1/2-BPS modes that
become massless (and then tachyonic) at (above) the corresponding
Hagedorn temperature.

Going to four dimensions and using techniques of $N_4=4$
supergravity, we were able to compute the exact effective potential
of all potential tachyonic modes, describing all three perturbative
instabilities of $N_4=4$ strings (heterotic, type IIA and type IIB)
simultaneously. We find that this potential has a global stable
minimum in a region where the heterotic string is weakly coupled, so
that the 6D string coupling $\lambda_H<(\sqrt{2}+1)/2$. At the
minimum, the temperature is fixed in terms of the heterotic string
tension (or in terms of the string coupling in type IIA units), the
four internal supercoordinates decouple, and the system is described
by a non-critical superstring in six dimensions. Supersymmetry,
although restored in perturbation theory, appears to be broken at the
non-perturbative level.

On the heterotic or type IIA side, the high-temperature limit
corresponds to a topological theory described by an $N_4=2$
supersymmetric sigma-model on a non-trivial hyper-K\"ahler manifold.
On the type IIB side, on the other hand, the high-temperature phase
corresponds to a tensionless string defined by a limit that
generalizes the large-N limit of Yang--Mills theory. It is very
interesting to study in detail the properties of these theories
describing the high-temperature phase of string theory, to generalize
these results to other compactifications with a lower number of
supersymmetries, and to study possible physical implications, e.g. in
cosmology as well as in the case of TeV strings.

\vspace{4mm}
\centerline{\bf Acknowledgements}
\noindent
The authors would like to thank L. Alvarez-Gaum\'e, S. Ferrara, E.
Kiritsis and A. Zaffaroni for valuable discussions. I. A. thanks the
CERN Theory Division, while J.-P. D. and C. K. thank the Centre de
Physique Th\'eorique of the Ecole Polytechnique for hospitality
during various stages of this research. This work was partially
supported by the EEC under the contracts TMR-ERBFMRX-CT96-0045 and
TMR-ERBFMRX-CT96-0090.
\vspace{4mm}

\section*{Appendix}
\renewcommand{\theequation}{A.\arabic{equation}}
\setcounter{equation}{0}

The $N=1$ supergravity scalar potential generated by the K\"ahler
function (\ref{Kis}) and the superpotential (\ref{Wis}) can be
written
\beq
\label{Apot1}
V =
\kappa^{-4}e^K\left(\Delta_S+\Delta_T+\Delta_U+
\Delta_++\Delta_-\right),
\eeq
with
\beq
\label{Apot2}
\begin{array}{rcl}
\Delta_S &=& |W-2sW_S|^2-|W|^2,
\crbig
\Delta_T &=& |W-2tW_T|^2-|W|^2,
\crbig
\Delta_U &=& |W-2uW_U|^2-|W|^2,
\crbig
\Delta_+ &=& |W|^2 + {1\over2}Y^+W_{A^+}(W_{A^+})^*
+2|Z^+|^2|W-W_{A^+}Z^+_A|^2 \crbig
&& + [1+Z^{+2}(Z^{+2})^*]|W_{A^+}Z_A^{+*}|^2
-|W-W_{A^+}Z^+_A-Z^{+2}W_{A^+}Z_A^{+*}|^2,
\crbig
\Delta_- &=& |W|^2 + {1\over2}Y^-W_{A^-}(W_{A^-})^*
+2|Z^-|^2|W-W_{A^-}Z^-_A|^2 \crbig
&& + [1+Z^{-2}(Z^{-2})^*]|W_{A^-}Z_A^{-*}|^2
-|W-W_{A^-}Z^-_A-Z^{-2}W_{A^-}Z_A^{-*}|^2.
\end{array}
\eeq
Each of the above quantities is a polynomial in the fields $S$, $T$,
$U$,
$Z_A^\pm$. The notation is
\beq\label{Apot3}
\begin{array}{rclrcl}
W_S &=& {\partial W\over\partial S}, \qquad &
W_T &=& {\partial W\over\partial T}, \crbig
W_U &=& {\partial W\over\partial U} , \qquad&
W_{A^\pm} &=& {\partial W\over\partial Z_A^\pm},
\end{array}
\eeq
repeated indices are summed over $A,B=1,2,3$,
$Y^\pm$ has been defined in Eq.(\ref{Yis}),
$$
|Z^\pm|^2 = Z_A^\pm Z_A^{\pm*}, \qquad\qquad Z^{\pm2} = Z_A^\pm
Z_A^\pm
$$
and $e^K=(8stuY^+Y^-)^{-1}$. Expressions (\ref{Apot1} and
(\ref{Apot2}) follow from the K\"ahler function $K$ only; the
structure of the superpotential has not been used. Notice also that
$V$ depends on {\it quadratic} combinations of the fields $Z_A^\pm$
and their conjugates. It is then invariant under
$Z_A^\pm\rightarrow-Z_A^\pm$ and stationary at $Z_A^\pm=0$ with
respect to these fields. Since $V(S,T,U,Z_A^\pm=0)\equiv0$,
$Z_A^\pm=0$ is a stable extremum for all values of $S$, $T$ and $U$.
We will in this appendix analyse this vacuum, which corresponds to
the low-temperature phase common to the three strings, compare the
scalar spectrum with masses of perturbative string states, and
identify the truncation relevant to the study of thermal
instabilities that is used in the body of the paper.

The calculation of the scalar mass matrices is a straightforward
exercise, using
$$
{\cal G}_{A^\pm}={\partial{\cal G}\over\partial Z_A^\pm}=0, \qquad
{\cal G}_S= -{1\over2s}, \qquad {\cal G}_T= -{1\over2t}, \qquad
{\cal G}_U= -{1\over2u},
$$
with ${\cal G}=K+\log|W|^2$. The mass matrix splits into four
sectors: $Z_1^\pm$ (heterotic winding modes), $Z_2^\pm$ (IIA
windings), $Z_3^\pm$ (IIB windings), $S,T,U$ (moduli). As already
mentioned, the moduli sector is trivially massless since the
potential at $Z_A^\pm=0$ is flat.

\vspace{4mm}\noindent {\bf 1) $Z_1^\pm$:}\hfill \\
In terms of the gravitino mass
\beq
\label{mgavapp}
m_{3/2}^2 = \kappa^{-2}e^{\cal G} = {1\over4\kappa^2stu}
= {1\over2\alpha^\prime_H tu} = {1\over2\alpha^\prime_{IIA} su}
= {1\over2\alpha^\prime_{IIB} st},
\eeq
the mass matrix in the $Z_1^\pm$ sector is:
\beq
\begin{array}{r}
Z_1^+: \crbig Z_1^-: \crbig Z_1^{+*}: \crbig Z_1^{-*} :
\end{array} \qquad
m^2_{3/2}\left(\begin{array}{cccc}
(tu-1)^2+2 & -2(tu-1) & -2 & -2(tu+1) \crbig
-2(tu-1) & (tu-1)^2+2 & -2(tu+1) & -2 \crbig
-2 & -2(tu+1) & (tu-1)^2+2 & -2(tu-1) \crbig
-2(tu+1) & -2 & -2(tu-1) & (tu-1)^2+2
\end{array}\right).
\eeq
Using mass formula (\ref{mass5}), the eigenstates and their masses
can be identified with (perturbative) heterotic states with momentum
and winding numbers $m$ and $n$:
\newpage
\vspace{4mm}
\begin{tabular}{llll}
${1\over2}\Re(Z_1^++Z_1^-):\quad$ & mass$^2=
{1\over2\alpha^\prime_H}\left[tu+(tu)^{-1}-6\right],\quad$ &
$m=\pm1,\quad$ & $n=\mp1$; \crbig
${1\over2}\Im(Z_1^+-Z_1^-):$ & mass$^2=
{1\over2\alpha^\prime_H}\left[tu+(tu)^{-1}-2\right],$ &
$m=0,$ & $n=\pm1$; \crbig
${1\over2}\Re(Z_1^+-Z_1^-):$ & mass$^2=
{1\over2\alpha^\prime_H}\left[tu+(tu)^{-1}+2\right],$ &
$m=0,$ & $n=\pm1$; \crbig
${1\over2}\Im(Z_1^++Z_1^-):$ & mass$^2=
{1\over2\alpha^\prime_H}\left[tu+9(tu)^{-1}-2\right],$ &
$m=\pm1,$ & $n=\pm1$.
\end{tabular}
\vspace{4mm}

\noindent
The first state is the lowest would-be tachyon at the origin of the
heterotic Hagedorn temperatures. The other three states cannot become
tachyonic and it is then sufficient to truncate the spectrum to
${1\over2}\Re(Z_1^++Z_1^-)$ to study the thermal instabilities
induced by perturbative heterotic states.

\vspace{4mm}\noindent {\bf 2) $Z_2^\pm$:}

\noindent
The mass matrix is:
$$
\begin{array}{r}
Z_2^+: \crbig Z_2^-: \crbig Z_2^{+*}: \crbig Z_2^{-*} :
\end{array} \qquad
m^2_{3/2}\left(\begin{array}{cccc}
(su)^2+2 & -2su & -2 & -2su \crbig
-2su & (su)^2+2 & -2su & -2 \crbig
-2 & -2su & (su)^2+2 & -2su \crbig
-2su & -2 & -2su & (su)^2+2
\end{array}\right).
$$
Again, using mass formula (\ref{mass5}), the eigenstates and their
masses can be identified with (perturbative) type IIA states with
momentum and winding numbers $m$ and $\tilde m^\prime$:

\vspace{4mm}
\begin{tabular}{llll}
${1\over2}\Re(Z_2^++Z_2^-):\quad$ & mass$^2=
{1\over2\alpha^\prime_{IIA}}\left[su-4\right],\quad$ &
$m=0,\,\,$ & $\tilde m^\prime=\pm1$; \crbig
${1\over2}\Im(Z_2^++Z_2^-):$ & mass$^2=
{1\over2\alpha^\prime_{IIA}}\left[su+4(su)^{-1}\right],$ &
$m=\pm1,$ & $\tilde m^\prime=\pm1$; \crbig
${1\over2}\Im(Z_2^+-Z_2^-):$ & mass$^2=
{1\over2\alpha^\prime_{IIA}}\left[su+4(su)^{-1}\right],$ &
$m=\pm1,$ & $\tilde m^\prime=\pm1$; \crbig
${1\over2}\Re(Z_2^+-Z_2^-):$ & mass$^2=
{1\over2\alpha^\prime_{IIA}}\left[su+4\right],$ &
$m=0,$ & $\tilde m^\prime=\pm1$.
\end{tabular}
\vspace{4mm}

\noindent
The first state is the lowest potential tachyon and one can truncate
the theory to this direction only for the study of thermal
instabilities induced by type IIA perturbative states.

\vspace{4mm}\noindent {\bf 3) $Z_3^\pm$:} \hfill

\noindent
The scalar mass matrix is obtained by replacing $u$ by $t$ in the
$Z_2^\pm$ mass matrix, as a result of IIA--IIB duality. The
discussion of the mass spectrum is then similar for string states
with momentum number $m$ and IIB winding $\tilde n^\prime$. Again,
thermal instabilities are generated in the field direction
${1\over2}\Re(Z_3^++Z_3^-)$ only.

In the body of the paper, we have used, in general, the scalar
potential truncated to directions $Z_A^+=Z_A^-=(Z_A^+)^*=(Z_A^-)^*$
only to enumerate the thermal phases of the theory. But we have also
checked by computing the complete mass matrices that this phase
structure is not modified by tachyons arising in other directions in
the scalar field space.

\end{document}